\documentclass[letterpaper,twocolumn,10pt]{article}
\usepackage{zhanggroup}

\usepackage{booktabs}
\usepackage{multirow}
\usepackage{xcolor}
\usepackage{xurl}
\usepackage{xspace}
\usepackage[T1]{fontenc}
\usepackage{graphicx}
\usepackage{amsmath,amsfonts}
\usepackage{algorithmic}
\usepackage{algorithm}

\newcommand{\mypara}[1]{\smallskip\noindent{\bf {#1}.}\xspace}
\newcommand{\method}{\textsc{BadBone}\xspace}
\newcommand{\revision}[1]{{{#1}}\xspace}

\begin{document}

\date{}

\title{\bf \method: Backdoor Attacks Against Backbone Models in\\ Visual Prompt Learning}

\author{
Ziqing Yang\textsuperscript{1}\ \ \
Rui Wen\textsuperscript{2}\ \ \
Xinlei He\textsuperscript{3}\ \ \
Yun Shen\textsuperscript{4}\ \ \
Michael Backes\textsuperscript{1}\ \ \
Yang Zhang\textsuperscript{1}\thanks{Corresponding author.}
\\
\\
\textsuperscript{1}\textit{CISPA Helmholtz Center for Information Security}\\
\textsuperscript{2}\textit{Institute of Science Tokyo} \ \ \ 
\textsuperscript{3}\textit{Wuhan University} \ \ \
\textsuperscript{4}\textit{Flexera}
}

\maketitle

\begin{abstract}
Prompt learning is a new machine learning paradigm that has attracted ample attention due to its simplicity and proven efficacy.
Despite its growing adoption, the security vulnerabilities associated with this paradigm remain underexplored.
In this work, we take the first step to propose \method, a stealthy and adaptive backdoor attack against prompt learning using bi-level optimization.
Instead of backdooring the prompt learning process, we aim to compromise a backbone model such that only target downstream tasks employing prompt learning inherit the backdoor vulnerability.
Extensive experiments on three different models and three datasets from various domains show that our targeted/untargeted backdoored models achieve high attack performance while maintaining utility on both pre-training and downstream tasks.
Moreover, we evaluate our approach against six state-of-the-art model-level defenses, including Neural Cleanse, ABS, MNTD, NAD, CLP, and D-BR.
The results demonstrate that these defenses are largely ineffective against our backdoored models and thus leave the effective defense as an important direction for future work.
Our code is available at \url{https://github.com/TrustAIRLab/BadBone}.
\end{abstract}

\section{Introduction}

Large-scale pre-trained models like ViT~\cite{DBKWZUDMHGUH21} and CLIP~\cite{RKHRGASAMCKS21} have demonstrated state-of-the-art results on computer vision tasks, owing to their capability of effectively capturing knowledge from massive data.
These models have been widely adopted as backbone models for downstream tasks~\cite{KTN20, MHB21, PWSCL21}.
Fine-tuning~\cite{DISFHS20, LCYLRBS20} is one of the most common approaches to transferring knowledge from backbone models to downstream tasks.
However, fine-tuning necessitates significant effort and computational resources to adjust the parameters of the backbone models.
Furthermore, the resulting fine-tuned models are inherently task-specific and lack the flexibility to be easily adapted to different tasks.

To address the limitations of fine-tuning, a new paradigm named \emph{prompt learning} has emerged~\cite{LYFJHN23, JTCCBHL22, BJSI22, HZDLS22}.
Given a backbone model and a downstream task, prompt learning aims to learn an input perturbation (i.e., a \emph{prompt}), which prompts the frozen backbone model to perform the downstream task via shifting the input data.
Compared to fine-tuning, prompt learning has several advantages.
First, prompt learning does not require updating the backbone model, which saves time and resources.
Second, a single backbone model can be applied to different downstream tasks by using different prompts, exemplifying that prompt learning is more versatile than fine-tuning.
According to previous research, prompt learning has already attained competitive performance in both computer vision~\cite{EGS19, TMCEVH21, JTCCBHL22, BJSI22, CYCZL22, ZLZHL22, BGDGE22, YSBZ23, LSPC23} and natural language processing (NLP) tasks~\cite{LL21, QE21, LYFJHN23} and is being commercially deployed, such as Landing AI~\cite{Landing_AI} and Amazon~\cite{KCSS23}.

\begin{figure}[!t]
\centering
\includegraphics[width=\columnwidth]{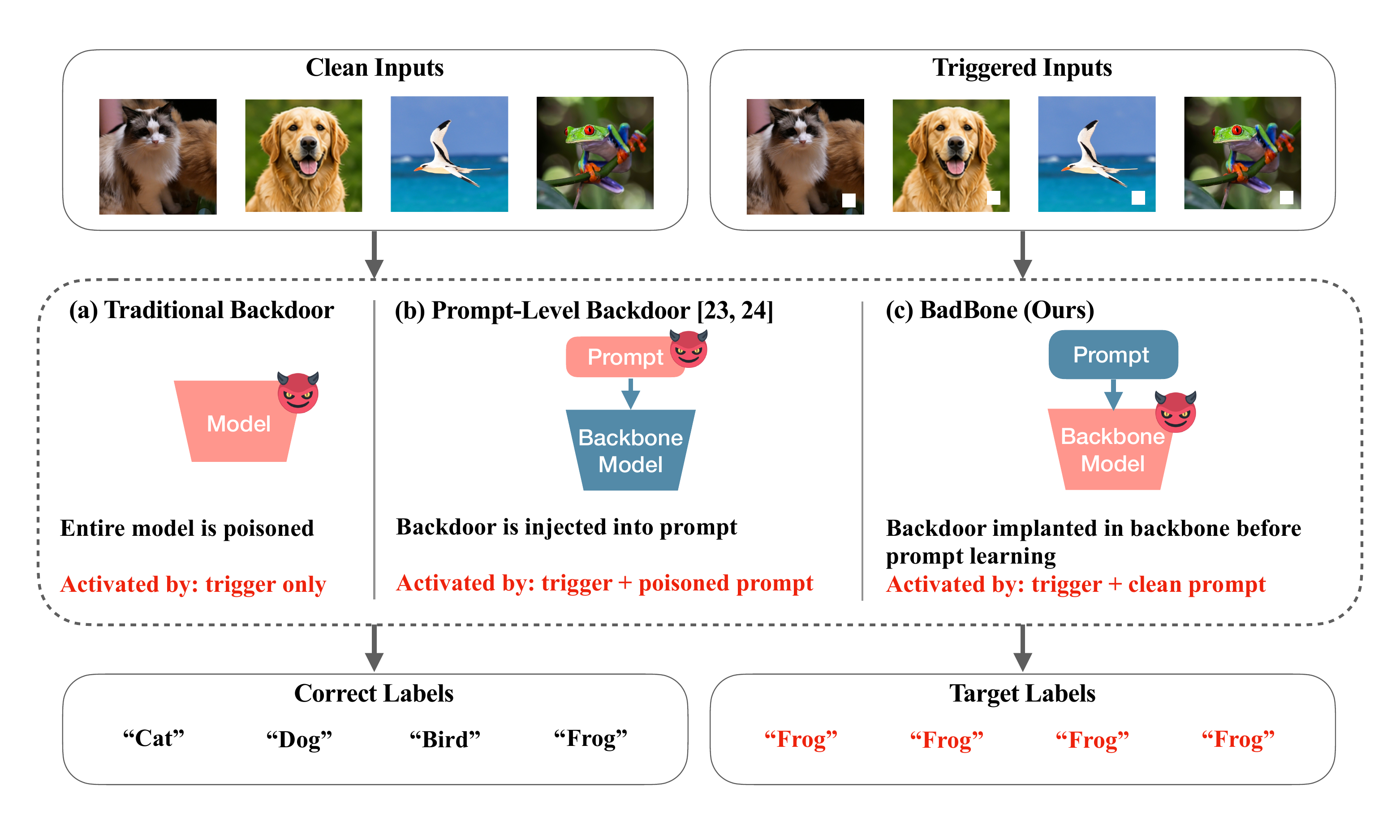}
\caption{
\revision{Comparison of three backdoor attack scenarios.
Compared to other backdoor methods, our \method has a stealthier activation mechanism and allows clean prompt learning.}
}
\label{figure:overview}
\end{figure}

Such a new learning paradigm, however, faces severe security risks, especially backdoor attacks~\cite{GDG17}.
Recall that a backbone model of prompt learning is frozen and thus can be \emph{reused} by downstream tasks.
One natural direction is implanting the backdoor into the prompt~\cite{DZLLW22, HZBSZ232}, i.e., a poisoned prompt is released to the victim without modifying the backbone model (see Figure~\ref{figure:overview}~(b)).
Despite being straightforward and practical, \emph{these attacks are limited to a single visual prompt by design.}
These attacks are rendered ineffective if the victim opts for a different visual prompt.
In contrast, we focus on backdooring the backbone model so that target downstream tasks employing prompt learning with this model will inherit the backdoor vulnerability, as illustrated in Figure~\ref{figure:overview}~(c).

\begin{figure*}[!t]
\centering
\includegraphics[width=0.9\textwidth]{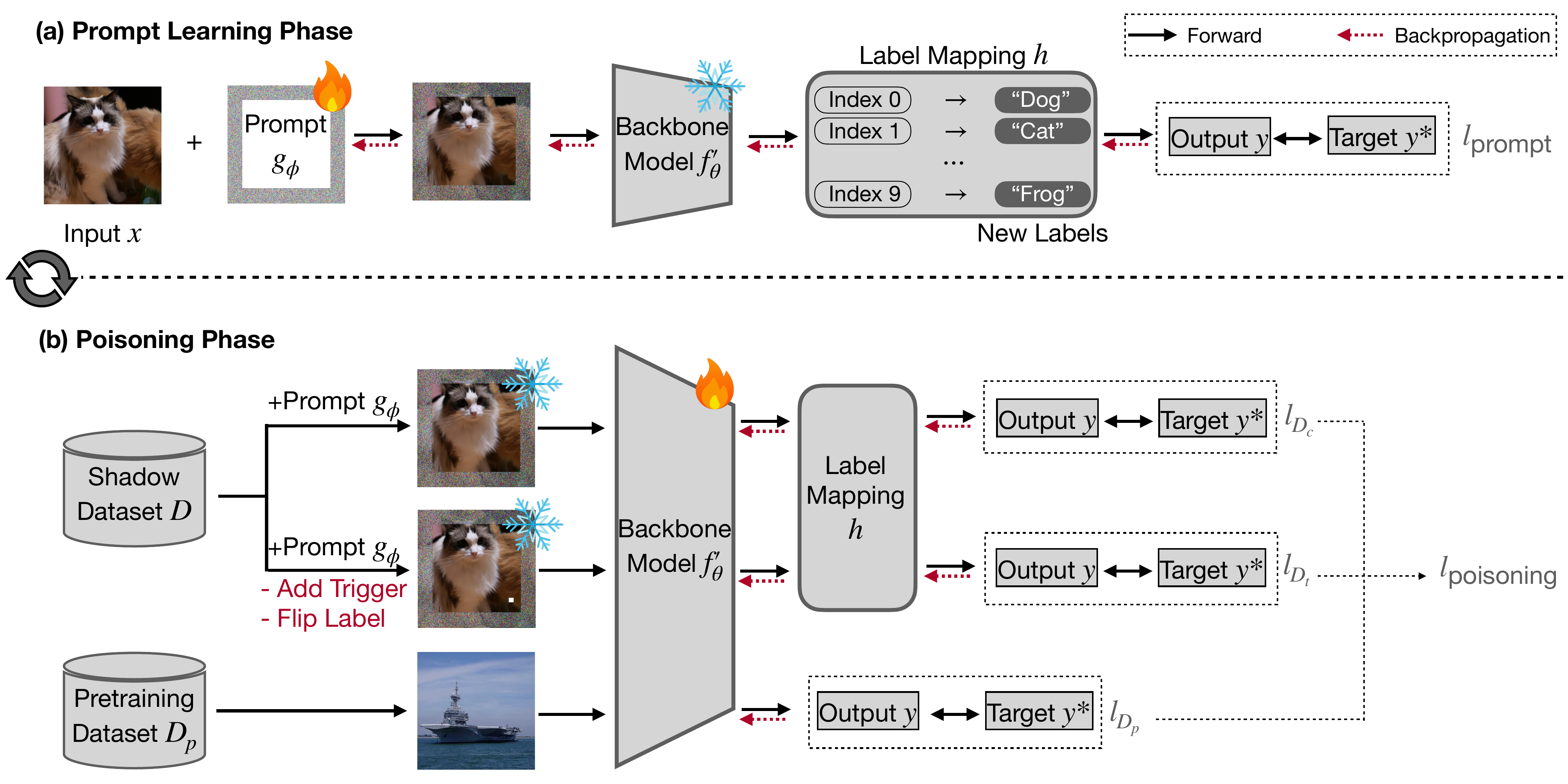}
\caption{
Framework of our \method, which iteratively learns a prompt (a) and poisons the backbone model with the poisoning dataset (b).
(a) Given an input image $x$, the prompt $g_{\phi}$ adds a pixel patch to it, which differs from the trigger in backdoor attacks.
Then, the prompted image is forwarded to the frozen backbone model $f_{\theta}$, and the label mapping function $h$ maps the output of the backbone model to downstream classes.
This phase aims to \emph{mock} to optimize $\phi$ for the downstream task.
(b) The backbone model $f_{\theta}$ is fine-tuned on the shadow dataset and the pretraining dataset while keeping the learned prompt frozen.
}
\label{figure:framework}
\end{figure*}

In this paper, we propose \method, the first backdoor attack targeting the backbone model in visual prompt learning based on bi-level optimization~\cite{D02}.
Our attack features a stealthy ``prompt-and-trigger co-activation'' mechanism, i.e., the backdoor remains dormant and difficult to detect until a learned prompt and a trigger appear together.
\revision{This is essential, as the attacker must ``anticipate'' how the victim will later learn prompts on downstream data, while simultaneously injecting a backdoor into the backbone.
A single-level or simpler optimization cannot capture this interaction between backbone poisoning and downstream prompt adaptation.}
As shown in Figure~\ref{figure:framework}, the adversary employs a framework that iteratively alternates between two stages: a prompt learning phase and a poisoning phase.
In the \emph{prompt learning phase}, the attacker mimics the victim's prompt generation process to ensure the backdoor is inherited by the downstream task.
In the \emph{poisoning phase}, the backdoor is injected into the backbone model.
Finally, the adversary can distribute the backdoored model online and provide the prompt learning instructions.
The backdoor is activated and exhibits malicious behavior only after an unsuspecting victim uses the model to generate their own prompt for a specific task.

We consider two different attack objectives, i.e., targeted and untargeted backdoor attacks.
For the \emph{targeted} attack, the adversary aims to misclassify all triggered samples to a target class, while the \emph{untargeted} attack is designed to misclassify all triggered samples, i.e., reducing the overall classification performance.
Note that our attacks succeed even if the victim conducts prompt learning using different forms of prompts other than the ones used to backdoor the backbone model.
Moreover, our attack only requires a shadow dataset with a similar distribution to the downstream dataset, rendering it a viable real-world attack.
This capability facilitates a more adaptable and transferable attack.

We evaluate our targeted and untargeted backdoor attacks on three widely used backbone models on three downstream image classification tasks.
In all cases, the results show that both backdoor attacks outperform the baseline.
For instance, on the CIFAR-10 dataset~\cite{CIFAR}, the targeted backdoor attack on the ResNet50 model~\cite{HZRS16} achieves a 98.66\% attack success rate, representing an 86.22\% performance gain over the baseline.
Experiments also show that the backdoored models maintain good utility on both the pre-training task and the downstream task (with a learned prompt).
Moreover, we examine the effectiveness of six model-level defenses developed to mitigate backdoor attacks.
Results show that most of our backdoored models can successfully bypass these defenses.

\begin{figure}[!t]
\centering
\includegraphics[width=\columnwidth]{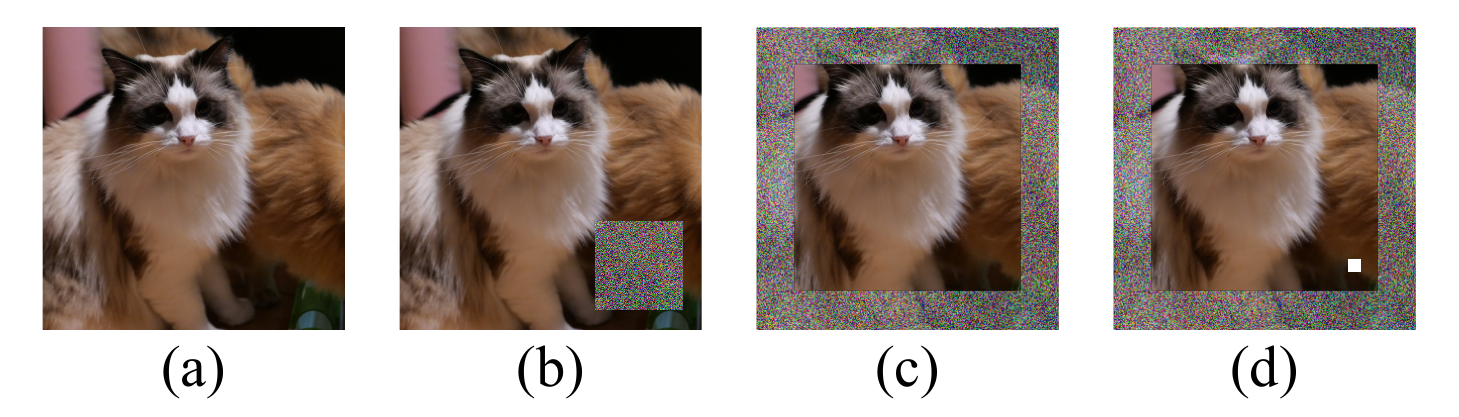}
\caption{Visual prompts can be visualized as adding a patch to the image.
(a) shows an input image $x$.
(b) and (c) are two different prompted formats of the image $g(x)$.
Specifically, (b) is a pixel patch, and (c) is padding around the image.
And (d) is a prompted image with a white square trigger at the bottom right corner.
}
\label{figure:prompt}
\end{figure}

Overall, our contributions are as follows:
\begin{itemize}
    \item We propose \method, the first targeted/untargeted backdoor attack against the backbone model in visual prompt learning.
    Our attack can implant backdoors into downstream tasks without compromising the prompt learning process.
    \item Empirical results demonstrate the effectiveness, stealthiness, and transferability of \method in various downstream tasks with different backbone models.
    \item We explore six defenses to mitigate our attacks.
    The results show that they cannot reliably defend against our attacks, highlighting the need to develop more advanced defenses for visual prompt learning.
\end{itemize}

\section{Preliminary}
\label{section:preliminary}

A \emph{backbone model} can be seen as a parameterized function $f_{\theta}: \mathcal{X} \to \mathbb{R}^M$, where $\theta$ represents the function's parameters and $\mathcal{X}$ denotes the input space.
Given a labeled downstream dataset $D=(\mathcal{X}, \mathcal{Y})$, a pre-trained backbone model $f_{\theta}: \mathcal{X}_{\text{pre}} \to \mathbb{R}^M$, and a pre-defined label mapping function $h: \mathbb{R}^M \to \mathbb{R}^N (M \geq N)$, \emph{visual prompt learning} learns a task-specific visual prompt $g_{\phi}: \mathcal{X} \to \mathcal{X}_{\text{pre}}$ that maps data in the input space of the downstream task $\mathcal{T}$ to the input space of the pre-training task $\mathcal{T}_{\text{pre}}$, parameterized by $\phi$.
$M$ and $N$ denote the number of classes in the pre-training and downstream tasks, respectively.
To link the relationship between labels in $\mathcal{T}$ and $\mathcal{T}_{\text{pre}}$, the \emph{mapping function $h$} is designed to map class indices of the downstream task to those of the pre-training task.
This mapping function discards unassigned (pre-training class) indices for loss computation~\cite{BJSI22}.
During the learning, the prompt $g_{\phi}$ is trained to maximize the likelihood of the correct label $y$,
\begin{equation}
\label{equation:prompt}
\begin{aligned}
& \max_{\phi} \mathbb{P}_{\theta, \phi}(y|h \circ f_{\theta} \circ g_{\phi} (x))\\
= & \max_{\phi} \mathbb{P}_{\theta, \phi}(y|F_{\theta, \phi}(x)),\ \ (x,y) \in D,
\end{aligned}
\end{equation}
where $\circ$ is the function composition operator and $F_{\theta, \phi}(\cdot)=h \circ f_{\theta} \circ g_{\phi}(\cdot)$ represents the \emph{prompted model}.
Note that, during the learning process, the gradient updates are only applied to the prompt parameters $\phi$, and the backbone model $f_{\theta}$ remains frozen.
Figure~\ref{figure:framework}~(a) illustrates a prompt learning process.

\mypara{Remark}
The learned \emph{prompt} can be visualized in the format of a patch in the pixel space and added to the image during inference, as shown in Figure~\ref{figure:prompt}.
In this work, we use the prompt of padding, as it consistently outperforms other design choices~\cite{BJSI22}.
Note that a \textit{patch} in this context differs from a \textit{trigger} employed in backdoor attacks, as illustrated in Figure~\ref{figure:prompt}~(d).
Instead, a patch serves as a visualization of a prompt that effectively instructs the backbone model to execute specific downstream tasks.

\section{Threat Model}
\label{section:threat_model}

\mypara{Attack Scenario}
We envision that the adversary is a malicious model provider who supplies a backdoored backbone model $f'$ to the victim.
The victim then generates a visual prompt $g'$ based on this model $f'$ for the target task $\mathcal{T}$, resulting in a backdoored task.
Our attack scenario is in line with the settings in recent literature~\cite{SRS17,JLG22,CT22,SJZLCSFYW21,STKP22}, where attackers may supply models to victims with limited ML expertise, e.g., medical research institutes, app developers, etc.
The adversary then gains knowledge of sampled downstream data $D$ of the target task $\mathcal{T}$.
We emphasize that the provider only supplies the (backdoored) backbone model (i.e., model weights) and allows victims to freely generate their own prompts.

\mypara{Attack Goals}
The adversary has two basic goals: utility and attack effectiveness.
For utility, the adversary must guarantee a competitive model utility of the backdoored backbone model $f'$ compared to the clean backbone model $f$.
From the victim's perspective, a high classification performance of $D^{\text{test}}_{\text{pre}}$ on the pre-training task $\mathcal{T}_{\text{pre}}$ is a key indicator that a backbone model encompasses good knowledge and can be transferred to different downstream tasks.
Additionally, prompt learning performance on clean samples from the target task $\mathcal{T}$ should be guaranteed.
Regarding effectiveness, in targeted attacks, all manipulated images should be misclassified as the target class $y_t$, while in untargeted attacks, all triggered images should be misclassified.
We further propose an additional goal, stealthiness.
The backdoor should be unlikely to be activated under inadequate conditions, i.e., the backdoor is activated only when both prompt learning and the trigger are present.
It should also be hard to detect, for example, be robust against backdoor detection and removal defenses.

\mypara{Attacker's Capabilities}
We assume that the adversary can train or customize an existing backbone model~\cite{JLG22}.
They can collect a shadow dataset $D$ similar to the victim's data for the target task $\mathcal{T}$~\cite{SJZLCSFYW21, STKP22, JLG22, LZLFY22}.
However, they cannot influence the prompt generation process.

\section{Methodology}
\label{section:methodology}

\subsection{Bi-level Optimization}
\label{section:bilevel}

\mypara{Objective}
The threat model indicates that our attack must account for the backbone model $f$, the visual prompt $g$, and their interaction. To address this challenge, we propose a bi-level optimization approach.
Specifically, we formulate an objective with ${\phi}$ and ${\theta}$ as the upper- and lower-level variables:
\begin{equation}
\begin{aligned}
\label{equation:bi-optimization}
& \phi^* = \arg \min_{\phi}{{\ell}_{\text{prompt}}(\theta^*, \phi)}, \\
& \text{s.t. }  \theta^* = \arg \min_{\theta}{{\ell}_{\text{poisoning}}(\theta, \phi)},
\end{aligned}
\end{equation}
where ${\ell}_{\text{poisoning}}$ and ${\ell}_{\text{prompt}}$ represent the loss terms for poisoning and prompt learning, respectively.
Intuitively, the first objective indicates that the prompt is optimized based on the backdoored backbone model, while the second shows how to inject the backdoor into the backbone model.

\mypara{Iterative Approximation}
Exactly solving Equation~\ref{equation:bi-optimization} is non-trivial.
Thus, we propose an iterative approach where we optimize $\phi$ and $\theta$ iteratively to approximate the solution.
Each iteration involves two phases: the prompt learning phase and the poisoning phase.
In the \emph{prompt learning phase}, the adversary freezes the backbone model $f_{\theta}$ and generates a prompt $g_{\phi}$ that mimics the behavior of the victim when generating a visual prompt on their own.
This prompt is generated based on the shadow dataset $D$ that shares a similar distribution with the data used by the victim for the target task.
In the \emph{poisoning phase}, we fix the prompt and poison the backbone model by fine-tuning it with our poisoning dataset.
This iterative approximation enables us to support both targeted and untargeted backdoor attacks elegantly.

\mypara{Note}
We stress that our goal is to implant the backdoor into the backbone model.
The prompt learning phase here is \emph{only} used for optimization purposes.
Our attack does \emph{not} interfere with the visual prompt generation process on the victim's side.

\subsection{Targeted Backdoor Attack}
\label{section:targeted_attack}

\mypara{Overview}
In targeted backdoor attacks, the adversary aims to misclassify the triggered images to a specific target label.
We outline the technical details of both phases below.

\mypara{Prompt Learning Phase}
In this phase, we simulate the process of prompt learning for the target task $\mathcal{T}$.
We freeze the backbone model $f_{\theta}$ and optimize a prompt $g_\phi$ using
${\ell}_{\text{prompt}}$, which is a categorical cross-entropy loss:
\begin{equation}
\label{equation:prompt_loss}
{\ell}_{\text{prompt}} = \mathbb{E}_{(x,y)\in D}{\sigma(F_{\theta, \phi}(x), y)},
\end{equation}
where $D$ is the shadow dataset and $\sigma(\cdot, \cdot)$ is the categorical cross-entropy loss.
$F_{\theta, \phi} (\cdot)$ represents the prompted model $h \circ f_{\theta} \circ g_{\phi}(\cdot)$, where $h$ is a pre-defined label mapping function.

\mypara{Poisoning Phase}
After obtaining the prompt $g_\phi$, we fix this prompt and poison the backbone model $f_{\theta}$ based on the poisoning dataset $D_{\text{poisoning}}$.
$D_{\text{poisoning}}$ consists of three subsets, i.e., a triggered dataset $D_t$ for the effectiveness goal, a clean dataset $D_c$ and a pre-training dataset $D_p$ for utility (see Section~\ref{section:threat_model}).
We outline the technical details below.
\begin{itemize}
\item \emph{Triggered dataset $D_t$} is constructed from a subset of the shadow dataset $D$.
We add a backdoor trigger on each image $x$ in this subset to construct triggered images:
\begin{equation}
\label{equation:add_trigger}
bd_{\text{trigger}}(x) = x \oplus \Delta_{\text{trigger}},
\end{equation}
where $\Delta_{\text{trigger}}$ denotes the trigger pattern defined by the attacker, and $\oplus$ represents adding the trigger on the image.
Then we flip their labels to a target label $y_t$:
\begin{equation}
\label{equation:targeted_flipping}
bd_{\text{flip}}(y) = y_t.
\end{equation}
\item \emph{Clean dataset $D_c$} is randomly sampled from the shadow dataset $D$.
It is used to maintain the performance on the target task $\mathcal{T}$ after prompt learning.
\item \emph{Pre-training dataset $D_p$} is randomly sampled from the original pre-training dataset that is used to train the backbone model $f$.
Note that the adversary is the malicious backbone model provider, hence having access to the original pre-training dataset.
\end{itemize}
Based on $D_{\text{poisoning}}$, we calculate the poisoning loss ${\ell}_{\text{poisoning}}$:
\begin{equation}
\begin{aligned}
\label{equation:poisoning_loss}
{\ell}_{\text{poisoning}} = \gamma * {\ell}_{D_t} + \beta * {\ell}_{D_c} + \alpha * {\ell}_{D_p},
\end{aligned}
\end{equation}
where $\gamma$, $\beta$, and $\alpha$ are the weights that balance the three key objectives: attack effectiveness (trigger loss ${\ell}_{D_t}$), downstream task utility (clean loss ${\ell}_{D_c}$), and pre-training task utility (pre-training loss ${\ell}_{D_p}$).
Each loss is a categorical cross-entropy loss (see Equation~\ref{equation:prompt_loss}) and is calculated based on the corresponding dataset (i.e., replacing $D$ with ${D_t}$, ${D_c}$, ${D_p}$, respectively).
The pseudo-code of \method can be found in Algorithm~\ref{algorithm:targeted}.

\begin{algorithm}[!t]
\caption{Algorithm of \method attack.}
\label{algorithm:targeted}
\begin{algorithmic}[1]
\STATE {\bfseries Data:} {A pre-trained backbone model $f_{\theta}$, a prompt $g_{\phi}$, a label mapping function $h$, a shadow dataset $D$, and a pre-training dataset $D_p$}
\STATE {\bfseries Input:} {Number of triggered samples $\#_{\text{trigger}}$, number of clean samples $\#_{\text{clean}}$, a backdoor label flipping strategy $bd_{\text{flip}}$, a backdoor trigger strategy $bd_{\text{trigger}}$, number of iterations $N_{\text{iters}}$, prompt learning epochs $E_p$, epochs $E_f$ for poisoning, learning rate of prompt learning $lr_{\text{prompt}}$, learning rate $lr_{\text{poisoning}}$ for poisoning, $\alpha$, $\beta$, and $\gamma$}
\STATE {\bfseries Output:} {Inject backdoor into the backbone model}
\STATE{// Construct triggered dataset $D_t$}

Randomly select $\#_{\text{trigger}}$ samples from $D$ as $D_t$;
\FOR{$(x,y)\in D_t$}{
    \STATE $x\leftarrow bd_{\text{trigger}}(x)$; {// Add triggers} \\
    \STATE $y\leftarrow bd_{\text{flip}}(y)$; {// Flip labels}
}
\ENDFOR
\STATE{// Construct clean dataset $D_c$}

Randomly select $\#_{\text{clean}}$ samples from $D$ as $D_c$;
\STATE{// Begin optimization}

\FOR{$l\leftarrow 1$ {\bfseries to} $N_{\text{iters}}$}{
      \STATE {// Prompt learning phase}

      Freeze $f_{\theta}$;
      Initialize $g_{\phi}$;
      \FOR{$e_p\leftarrow 1$ {\bfseries to} $E_p$}{
            \STATE Computing ${\ell}_{\text{prompt}}$ (Equation~\ref{equation:prompt_loss})\;
            \STATE Update $\phi \leftarrow \phi - lr_{\text{prompt}} * \nabla {\ell}_{\text{prompt}}$\;
      }
      \ENDFOR
      \STATE{// Poisoning phase}

      Freeze $g_{\phi}$;
      \FOR{$e_f\leftarrow 1$ {\bfseries to} $E_f$}{
            \STATE Compute ${\ell}_{\text{poisoning}}$ (Equation~\ref{equation:poisoning_loss})\;
            \STATE Update $\theta \leftarrow \theta - lr_{\text{poisoning}} * \nabla {\ell}_{\text{poisoning}}$\;
      }
      \ENDFOR
}
\ENDFOR
\end{algorithmic}
\end{algorithm}

\subsection{Untargeted Backdoor Attack}

\mypara{Overview}
The goal of the untargeted attack is to misclassify triggered images in general.
We follow the framework of the targeted backdoor attack, where we iteratively conduct prompt learning and optimize the backbone model.
We reuse the same prompt learning phase and revise the poisoning phase, which is introduced below.

\mypara{Poisoning Phase}
The only difference from the targeted attack is the construction of the triggered dataset $D_t$.
We first select a subset from the shadow dataset $D$ and add a backdoor trigger on each image in the subset.
Instead of assigning to one specific target label, we employ two strategies for labeling triggered images, i.e., \emph{next flipping} and \emph{random flipping}:
\begin{equation}
\label{equation:next_flipping}
bd_{\text{flip}}(y) = y+1 \mod N,
\end{equation}
\begin{equation}
\label{equation:random_flipping}
bd_{\text{flip}}(y) = \text{random}(\mathcal{Y} \setminus \{y\}),
\end{equation}
where $N$ denotes the number of target task classes and $\mathcal{Y}$ is the output class set.
For the next flipping (Equation~\ref{equation:next_flipping}), we flip the label of each triggered image $y$ to the next class (i.e., $y+1 \mod N$).
For random flipping (Equation~\ref{equation:random_flipping}), the label of each triggered image is randomly flipped to a class other than its original class.
With the triggered dataset $D_t$, we use the same trigger loss ${\ell}_{D_t}$ for our untargeted backdoor attack, where the model tends to misclassify the triggered images so that the effectiveness goal of the untargeted backdoor attack can be achieved.
The poisoning loss follows Equation~\ref{equation:poisoning_loss}.

\subsection{Novelty}
\label{section:methodology_discussion}

Instead of injecting backdoors into the prompt during prompt learning~\cite{DZLLW22, HZBSZ232}, our work targets the backbone model.
Existing backdoor attacks launched on backbone models~\cite{JLG22, SJZLCSFYW21}, however, focus on different paradigms.
Specifically, Jia et al.~\cite{JLG22} focus on self-supervised models, crafting an image encoder that produces similar feature vectors for the triggered images and images from the target class.
Similarly, Shen et al.~\cite{SJZLCSFYW21} use predefined output representation to backdoor pre-trained language models.
In contrast, we focus on prompt learning and thus allow users to create their own visual prompts for the target task $\mathcal{T}$, showing the adaptability of our \method.
This also requires addressing the interaction among the visual prompt, trigger, and backbone model simultaneously.
The complexity renders the approach from Jia et al.~\cite{JLG22} inapplicable as it optimizes the image encoder in a standalone manner.
Besides, the backdoor of our \method is activated only with both prompt learning and the trigger.
Such a prompt-and-trigger co-activation mechanism ensures the stealthiness of our attack, which is further validated empirically in Section~\ref{section:effective}.

\section{Experiments}
\label{section:experiments}

\subsection{Experimental Settings}

\begin{table*}[t!]
\centering
\caption{Dataset statistics}
\label{table:dataset}
\setlength{\tabcolsep}{3pt}
\begin{tabular}{l c c c c c l}
\toprule
Dataset & $\#_{\text{shadow}}^{\text{train}}$ & $\#_{\text{victim}}^{\text{train}}$ & $\#_{\text{shadow}}^{\text{test}}$ & $\#_{\text{victim}}^{\text{test}}$ & Classes & Category \\
\midrule
{CIFAR-10} & 25,000 & 25,000 & 5,000 & 5,000 & 10 & Nature \\
{SVHN} & 36,629 & 36,628 & 13,016 & 13,016 & 10 & Digits \\
{EuroSAT} & 11,000 & 11,000 & 2,500 & 2,500 & 10 & Satellite \\
\bottomrule
\end{tabular}
\end{table*}

\mypara{Datasets}
\revision{We consider four benchmark datasets that are widely used in computer vision tasks, including visual prompt learning~\cite{JTCCBHL22, BJSI22}.
Their respective details are outlined below.}

\begin{itemize}
\item \textbf{ImageNet-1k~\cite{DDSLLF09}}.
ImageNet-1k spans 1,000 object classes and contains 1,281,167 training and 50,000 validation colored images.
Each image has a size of 224$\times$224.
In the experiments, we form a subset of ImageNet-1k by randomly selecting 50 images from each class from the training dataset.
We denote this subset as ImageNet-1k~(50k) and use it to calculate the pre-training loss in our attack.
We evaluate the performance of the pre-training task, i.e., the model utility, on the original validation split of ImageNet-1k.

\item \textbf{CIFAR-10~\cite{CIFAR}}.
CIFAR-10 consists of 60,000 colored images, which are equally distributed over the following ten classes: airplane, automobile, bird, cat, deer, dog, frog, horse, ship, and truck.
Each image has a size of 32$\times$32.
There are 50,000 images and 10,000 images in the training and test datasets, respectively.

\item \textbf{SVHN~\cite{NWCBWN11}}.
SVHN is a digit classification dataset obtained from house numbers in Google Street View.\footnote{\url{https://www.google.com/streetview/}.}
It contains 600,000 colored images of printed digits cropped from pictures of house number plates distributed across ten classes, i.e., from 0 to 9.
Each image has a size of 32$\times$32.
Following the officially provided data splits, there are respectively 73,257 and 26,032 images in the training and test datasets.

\item \textbf{EuroSAT~\cite{HBDB19}}.
EuroSAT is a benchmark dataset for land use and land cover classification.
It is based on Sentinel-2 satellite images covering 13 spectral bands and consists of 27,000 colored images labeled with ten classes: forest, annual crop, highway, herbaceous vegetation, pasture, residential, river, industrial, permanent crop, and sea/lake.
Each image has a size of 64$\times$64.
We randomly select 22,000 and 5,000 images as the training and test datasets.
\end{itemize}

We randomly select 50\% of the images from the training split of the above downstream datasets, i.e., CIFAR-10, SVHN, and EuroSAT, to form our shadow dataset $D$.
The remaining 50\% images serve as the training datasets used by the victim.
Statistics of these datasets are summarized in Table~\ref{table:dataset}.

\mypara{Target Backbone Models}
We focus on image classification tasks, a common application for exploring backdoor attacks.
In our experiments, we select three backbone models, in which ResNet18~\cite{HZRS16} and ResNet50~\cite{HZRS16} are obtained from PyTorch's official Torchvision library,\footnote{\url{https://github.com/pytorch/vision}.} and Big Transfer (BiT-M-RN50)~\cite{KBZPYGH20} is from PyTorch Image Models.\footnote{\url{https://github.com/rwightman/pytorch-image-models}.}
All these models have pre-trained on the official training split of ImageNet-1k~\cite{DDSLLF09}.
More specifically, ResNet18 and ResNet50 are trained on the official training split of ImageNet-1k~\cite{DDSLLF09}.
BiT-M-RN50 is trained on a larger dataset ImageNet-21k~\cite{DDSLLF09} and fine-tuned on the official training split of ImageNet-1k.\footnote{
Since ImageNet-21k has no official test dataset, Kolesnikov et al.~\cite{KBZPYGH20} fine-tuned the model on ImageNet-1k and evaluated it on the ImageNet-1k test dataset.
We follow their settings and regard the image classification on ImageNet-1k as the pre-training task for BiT-M-RN50.}

\mypara{Configuration}
We use ImageNet-1k~(50k) as $D_p$ (pre-training dataset) for all experiments.
We randomly sample 5,000 images ($\#_{\text{clean}}$) from the shadow dataset $D$ as $D_c$ (clean dataset) and other 5,000 images ($\#_{\text{trigger}}$) from $D$ to construct the triggered dataset $D_t$.
The trigger is fixed to a 10$\times$10 white patch at the bottom right corner, which is relatively small compared with the input size 224$\times$224 (see Figure~\ref{figure:prompt}~(d)).
For targeted backdoor attacks, we map the triggered images to ``automobile'' in CIFAR-10, digit ``1'' in SVHN, and ``annual crop'' in EuroSAT.
For untargeted backdoor attacks, we flip the label of the triggered images using the next flipping strategy, e.g., class ``2'' $\rightarrow$ class ``3.''

\mypara{Hyper-Parameters}
The default hyper-parameters used in our experiments are as follows.
$N_{\text{iters}}$ is set to 2 (see Section~\ref{section:ablation_study} for additional details).
In the prompt learning phase, we run 100 epochs with a learning rate $lr_{\text{prompt}}$ of 40 in the first iteration and adjust to run only 10 epochs with $lr_{\text{prompt}}=$0.001 in the succeeding iterations.
The batch size is fixed to 128 for all the models in this phase.
Following the settings proposed by Bahng et al.~\cite{BJSI22}, we resize all images to 224$\times$224 to match the input size of backbone models and fix the padding size of the visual prompt to 30.
We use a hard-coded label mapping method~\cite{EGS19} and arbitrarily map downstream class indices to pre-training class indices, discarding unassigned indices for loss computation.
We compute the cross-entropy loss over the downstream class indices.
In the poisoning phase, we empirically set the hyper-parameters in the loss function ($\alpha, \beta, \gamma$) to 1 and optimize the backbone model for 10 epochs with a learning rate $lr_{\text{poisoning}}$ of 0.001.
The batch size is 128 for loading the pre-training dataset.
The batch size of the poisoning dataset is dynamically adjusted based on the batch number of the pre-training dataset, i.e., dividing the total number of samples by the batch number and rounding down.
All the learning process is optimized by SGD decayed with a cosine scheduler~\cite{LH17}.

\mypara{Evaluation Metrics}
We consider three evaluation metrics.
\begin{itemize}
\item \textbf{Classification Accuracy (\textit{Acc})} evaluates the performance on the pre-training task ($\textit{Acc}_{\text{pre}}$) on the backbone model $f_{\theta}$ and the target task ($\textit{Acc}_{\text{target}}$) on the prompted model $F_{\theta, \phi}$.
$\textit{Acc}_{\text{pre}}$ assesses if a backdoored backbone model $f'$ has a similar or better performance than a clean backbone model, while $\textit{Acc}_{\text{target}}$ assesses if the model $f'$ can perform well on the clean test data.
The higher $\textit{Acc}_{\text{pre}}$ and $\textit{Acc}_{\text{target}}$ are, the better the model utility is.
\item \textbf{Attack Success Rate (\textit{ASR})} measures the attack effectiveness of the backdoored model on a triggered test dataset for the targeted backdoor attack.
The higher the \textit{ASR} is, the better the targeted backdoor attack performance is (the effectiveness goal mentioned in Section~\ref{section:threat_model}).
\item \textbf{Misclassification Rate (\textit{MR})} measures the average misclassification rate for classifying triggered images to their original label.
It indicates how the model misclassifies images when they contain a trigger.
We exclusively use this metric to measure the effectiveness of the untargeted backdoor attack.
The higher the \textit{MR} is, the better the untargeted backdoor attack performs.
\end{itemize}

\begin{table*}[!t]
\centering
\caption{Performance of our \method (\%).
The numbers shown in the brackets are the baselines.
\textit{ASR/MR} baselines are the performance of the clean prompted models on the triggered images.
$\textit{Acc}_{\text{target}}$ and $\textit{Acc}_{\text{pre}}$ baselines are the accuracy of the clean model on the target and pre-training task, respectively.
}
\label{table:performance_all}
\setlength{\tabcolsep}{3pt}
\begin{tabular}{lcccc|ccc}
\toprule
\multirow{2}{*}{Dataset} & \multirow{2}{*}{Model} & \multicolumn{3}{c}{Targeted} & \multicolumn{3}{c}{Untargeted} \\
& & \textit{ASR} & $\textit{Acc}_{\text{target}}$ & $\textit{Acc}_{\text{pre}}$ & \textit{MR} & $\textit{Acc}_{\text{target}}$ & $\textit{Acc}_{\text{pre}}$\\
\midrule
\multirow{3}{*}{CIFAR-10} & ResNet18 & 97.64 (17.02) & 85.88 (54.10) & 66.12 (69.76) & 92.38 (50.88) & 86.14 (54.10) & 65.93 (69.76) \\
 & ResNet50 & 98.66 (12.44) & 91.04 (51.58) & 72.03 (76.15) & 94.88 (49.20) & 91.92 (51.58) & 71.81 (76.15) \\
 & BiT-M-RN50 & 98.92 (11.52) & 93.38 (61.66) & 72.11 (74.02) & 97.74 (41.26) & 94.02 (61.66) & 71.94 (74.02) \\
\midrule
\multirow{3}{*}{SVHN} & ResNet18 & 98.43 (25.51)& 92.03 (61.39) & 66.38 (69.76) & 96.16 (39.77) & 92.44 (61.39) & 66.19 (69.76) \\
 & ResNet50 & 99.19 (22.96) & 93.16 (57.91) & 71.90 (76.15) & 97.30 (42.51) & 94.10 (57.91) & 71.91 (76.15) \\
 & BiT-M-RN50 & 99.51 (20.00) & 91.56 (73.15) & 72.38 (74.02) & 97.80 (29.24) & 94.12 (73.15) & 71.95 (74.02) \\
\midrule
\multirow{3}{*}{EuroSAT} & ResNet18 & 94.72 (11.68) & 94.52 (76.68) & 66.46 (69.76) & 93.16 (26.68) & 95.20 (76.68) & 66.18 (69.76) \\
 & ResNet50 & 96.92 (10.68) & 95.48 (75.68) & 72.03 (76.15) & 95.52 (25.16) & 96.20 (75.68) & 71.86 (76.15) \\
 & BiT-M-RN50 & 98.56 (10.44) & 95.84 (84.84) & 72.65 (74.02) & 96.96 (23.72) & 96.92 (84.84) & 72.34 (74.02) \\
\bottomrule
\end{tabular}
\end{table*}

\subsection{Attack Performance}
\label{section:effective}

Table~\ref{table:performance_all} shows the performance of our targeted and untargeted attacks on three downstream datasets and three different models.

\mypara{Effectiveness}
\textit{ASR/MR} baselines (numbers in brackets) evaluate the triggered images on the clean prompted models, i.e., after conducting prompt learning on the clean backbone model.
Based on the results, our backdoor attack significantly outperforms clean models.
For example, the adversary achieves a 98.92\% \textit{ASR} with the backdoored BiT-M-RN50 model on the CIFAR-10 dataset, which is an 87.40\% performance gain over the baseline.
The untargeted backdoored ResNet18 model on EuroSAT also achieves a 93.16\% \textit{MR}, outperforming the baseline by 66.48\%.
Our results exemplify the effectiveness of our backdoor attack.

\mypara{Utility}
$\textit{Acc}_{\text{target}}$ and $\textit{Acc}_{\text{pre}}$ baselines (numbers in brackets) measure the accuracy of the clean model on the target and pre-training tasks, respectively.
Results show that the utility of both backbone and prompted models after our attack is well-preserved.
Take the untargeted backdoored ResNet18 model on EuroSAT as an example.
Its $\textit{Acc}_{\text{pre}}$ achieves 66.18\% and its $\textit{Acc}_{\text{target}}$ reaches 95.20\%, both of which are comparable with the baselines.
Note that with a generated prompt, the $\textit{Acc}_{\text{target}}$ represents an 18.52\% performance gain over the baseline.
The reason for this improved performance is that the backbone model learned the target task's knowledge from the clean dataset $D_c$ during the attack process.

\mypara{Stealthiness}
A key aspect of our attack's stealthiness is that it is a prompt-and-trigger co-activation mechanism.
In other words, given a backdoored backbone model, the backdoor is activated only with both prompt learning and the trigger.
This makes the backdoor difficult to detect, as it remains dormant under normal conditions.
\begin{itemize}
\item \textbf{Prompt Alone is Insufficient.}
The prompted model exhibits no backdoor behavior without a trigger.
This is evidenced by its high accuracy on clean test images, which is comparable to, and in some cases exceeds, the clean model's performance.
For instance, the backdoored BiT-M-RN50 model achieves 93.38\% accuracy on the clean CIFAR-10 test set.
\item \textbf{Trigger Alone is Insufficient.}
\revision{
Likewise, the trigger alone fails to activate the backdoor.
We evaluate the $\textit{ASR}/\textit{MR}$ of the triggered images on the backbone model for the pre-training task.
Specifically, we first add triggers to each image in the official test split of the ImageNet dataset, which is the pre-training dataset for the three models (ResNet18, ResNet50, and BiT-M-RN50).
Then we evaluate the $\textit{ASR}/\textit{MR}$ of the triggered images on the backdoored backbone model, without equipping it with a prompt.
Under such a circumstance, $\textit{ASR}$ becomes measuring whether the triggered images will be classified as the target label index, i.e., index ``1,'' while $\textit{MR}$ is the mis-classification rate on the pre-training task.
Results shown in Table~\ref{table:trigger_alone} indicate that the backdoored backbone model is indistinguishable from a clean one when tested with triggered inputs alone.
When evaluating on the pre-training task, the $ASR$ for the CIFAR-10-backdoored model was a mere 0.10\%, identical to the $ASR$ of a clean, un-poisoned model under the same conditions.
This result confirms that the backdoor remains dormant, indicating that the malicious behavior is only activated when both the trigger and a learned downstream prompt are present.}
\end{itemize}

\begin{table}[!t]
\centering
\caption{
\revision{Stealthiness of our \method under the trigger alone setting.
$\textit{ASR}/\textit{MR}$ is evaluated on the CIFAR-10-backdoored or clean backbone model with the triggered images in the \emph{pre-training} task, which differs from the main experiments.}
}
\label{table:trigger_alone}
\setlength{\tabcolsep}{3pt}
\begin{tabular}{llcc}
\toprule
 {Model} & Status & $\textit{ASR}$ & $\textit{MR}$\\
\midrule
\multirow{2}{*}{ResNet18} & Clean Backbone & 0.10 & 30.62 \\
& Backdoored Backbone & 0.10 & 30.62 \\
\midrule
\multirow{2}{*}{ResNet50} & Clean Backbone & 0.10 & 24.16 \\
& Backdoored Backbone & 0.10 & 24.16 \\
\midrule
\multirow{2}{*}{BiT-M-RN50} & Clean Backbone & 0.11 & 26.44 \\
& Backdoored Backbone & 0.11 & 26.44 \\
\bottomrule
\end{tabular}
\end{table}

\begin{table}[t!]
\centering
\caption{
\revision{Average runtime (GPU hours) per epoch in each phase.}
}
\label{table:runtime}
\setlength{\tabcolsep}{3pt}
\begin{tabular}{l c c}
\toprule
Phase & Runtime & $\#$Epochs per Loop \\
\midrule
Prompt Learning & 29.23s & 10 or 100 \\
Poisoning & 1912.19s & 10 \\
\bottomrule
\end{tabular}
\end{table}

\mypara{Computational Cost}
\revision{
As \method involves bi-level optimization, we report both memory usage and runtime (in GPU hours) for the target backdoor attack.
We conduct experiments on a ResNet18 model trained on CIFAR10 using an NVIDIA A100 GPU (80 GB).
As shown in Table~\ref{table:runtime}, each epoch in the prompt learning phase requires approximately 30 seconds, whereas each epoch in the poisoning phase takes around 32 minutes.
This indicates that prompt learning is computationally efficient, while the primary cost arises from backbone fine-tuning.
Importantly, this cost is expected and reasonable, as fine-tuning the backbone over multiple epochs is a standard practice in existing backdoor attack methods.
Therefore, our computational overhead is in line with prior work rather than introducing additional burden.
For the full training process with two iterations, the total runtime is approximately 11.5 GPU hours.
In terms of memory, \method reaches a peak GPU usage of 15.51 GB, with a steady-state footprint of around 15.50 GB.
This corresponds to an incremental allocation of approximately 14.97 GB above the warm idle baseline.
}

\mypara{Summary}
We show that the backdoored backbone model enables the adversary to achieve good attack performance in targeted and untargeted attacks.
At the same time, both attacks are stealthy and can guarantee a competitive model utility on both the pre-training task and the target task.

\subsection{Attack Transferability}
\label{section:transfer}

Our attack also exhibits high transferability, allowing different prompt designs and out-of-distribution datasets.
Here, we employ the ResNet18 model with the CIFAR-10 dataset.

\begin{figure}[!t]
\centering
\subfloat[\footnotesize{Targeted}]{\includegraphics[width=\columnwidth]{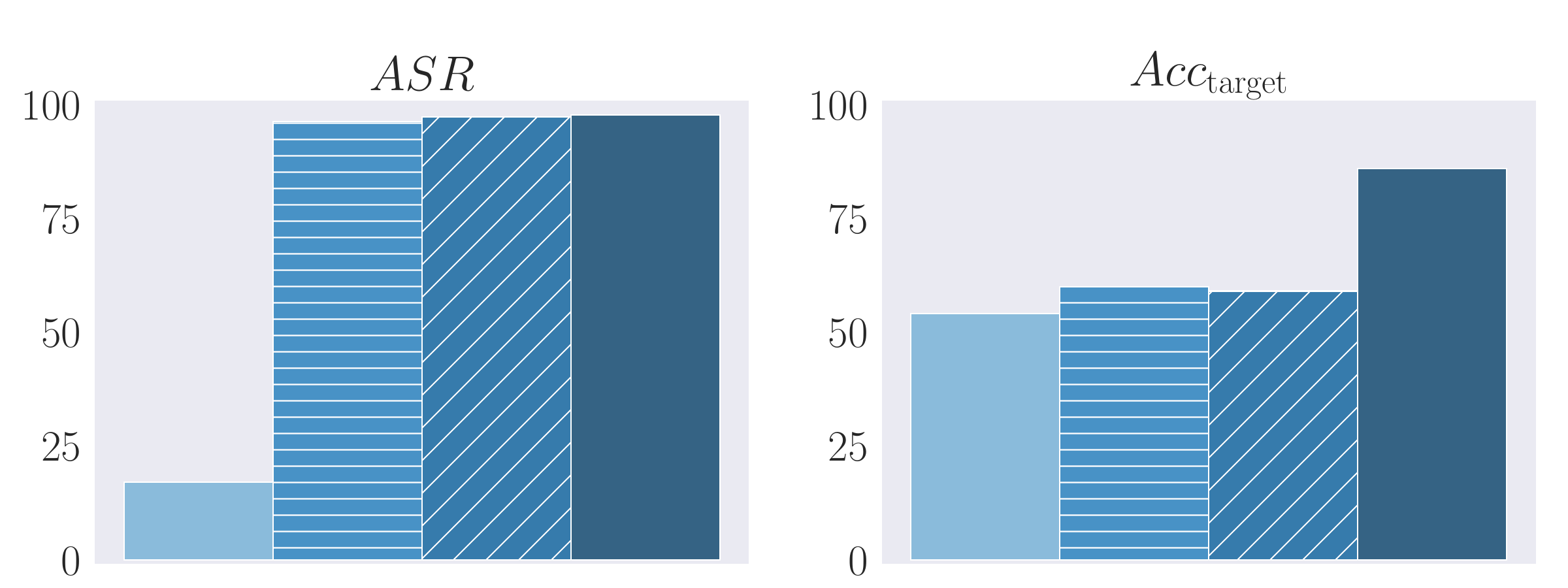}%
\label{figure:prompt_design_target}}
\hfil
\subfloat[\footnotesize{Untargeted}]{\includegraphics[width=\columnwidth]{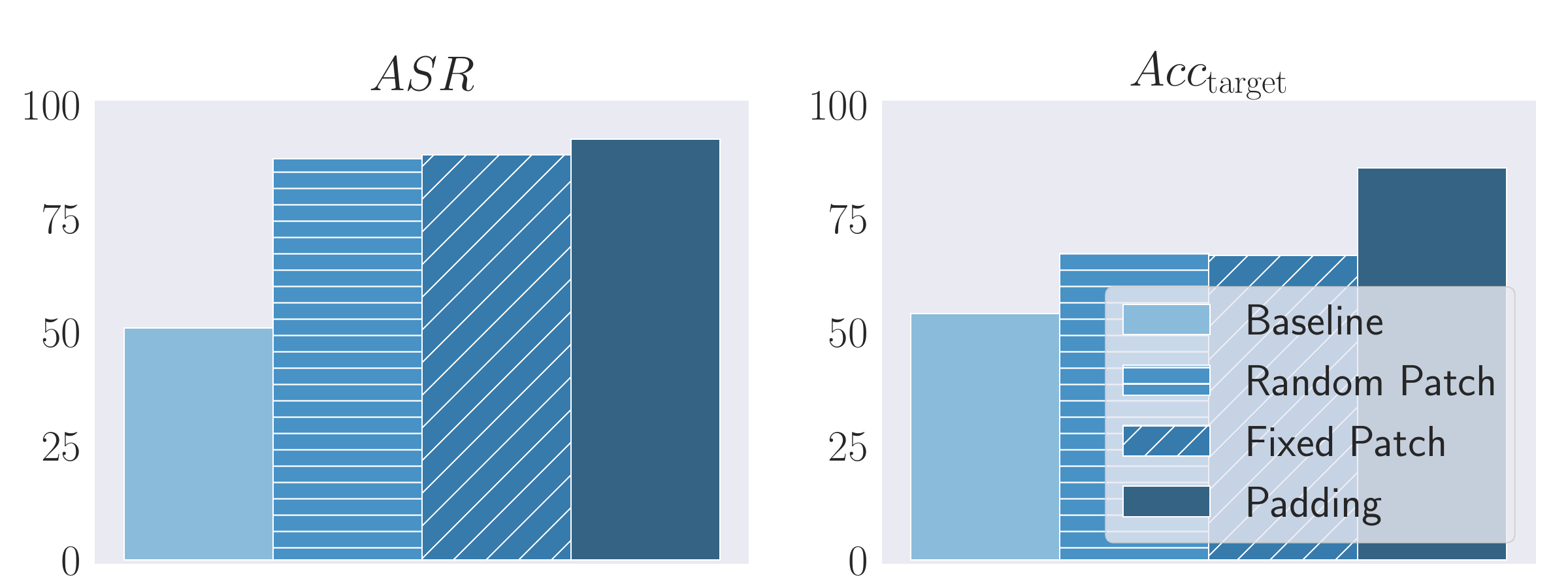}%
\label{figure:prompt_design_untarget}}
\caption{Influence of the prompt designs.
}
\label{figure:prompt_design}
\end{figure}

\mypara{Transfer to Different Prompt Designs}
In the previous experiments, we use the padding prompt (see Figure~\ref{figure:prompt}~(c)) in the prompt learning process.
We expect that, even if the victim conducts prompt learning using different forms of prompt, the adversarial goal is achieved.
Using the same backdoored backbone model, we evaluate the attack performance using a different form of prompt, e.g., a fixed/random patch on the image (see Figure~\ref{figure:prompt}~(b)).
The patch size remains the same, i.e., 30.
Note that the backdoored backbone remains frozen; thus, the $\textit{Acc}_{\text{pre}}$ will not change.
Based on Figure~\ref{figure:prompt_design}, we observe that the adversarial goal can still be achieved with different forms of prompts.
For example, the adversary achieves a 97.28\% \textit{ASR} when using a fixed patch.
The results also indicate that using different forms of prompts maintains utility well.
The $\textit{Acc}_{\text{target}}$ of random/fixed patch is higher than that of the baseline.

\mypara{Transfer to Different Datasets}
Domain shift/adaptation remains an ongoing research topic in ML~\cite{PY09}.
In our experiments, the shadow dataset has the same distribution as the downstream dataset used by the victim.
We relax this assumption by using a shadow dataset with a similar distribution to the downstream dataset to evaluate our attack's generalizability.
We thus introduce STL-10~\cite{CNL11}, which contains 13,000 96$\times$96 natural images, among which 5,000 images are partitioned for training while the remaining 8,000 images are for testing.
For our experiment, we use STL-10 as our shadow dataset $D$ to craft the backdoored backbone model and evaluate the target task on CIFAR-10.
For the labels in the shadow dataset $D$, we re-order the class indices of STL-10 to align with CIFAR-10 class indices.
To build the poisoning dataset, we randomly select 2,000 images from STL-10 as the clean dataset $D_c$ and another 2,000 images to form the triggered dataset $D_t$.
In the targeted attack, we map the label of the triggered images to class 1 (the ``car'' in STL-10), which corresponds to the target class ``automobile'' in CIFAR-10.
In the prompt learning phase, we run 100 epochs with a learning rate $lr_{\text{prompt}}$ of 100 in the first iteration and 20 epochs with $lr_{\text{prompt}}=$0.01 in the succeeding iterations.
Experiments show that STL-10 requires a larger learning rate in prompt learning than CIFAR-10, which indicates that they have different distributions.
In the poisoning phase, we optimize the backbone model for 10 epochs with a learning rate $lr_{\text{poisoning}}$ of 0.001 in every iteration.
Note that $lr_{\text{prompt}}$ is slightly different so that we can adapt to various datasets and backbone models.
However, $lr_{\text{poisoning}}$ remains fixed at 0.001 throughout the paper.
With the poisoned backbone model, we conduct prompt learning using CIFAR-10 as the downstream dataset and measure the attack performance on the prompted model.
The learning rate used in the evaluation is 0.1.
Results in Table~\ref{table:stl10} show that using STL-10 as the shadow dataset for backdooring the target task on the CIFAR-10 dataset can also attain effectiveness and utility goals.
For example, the \textit{ASR} of the targeted backdoored model is 97.64\%, which has an 80.62\% gain over the baseline.
The $\textit{Acc}_{\text{pre}}$ of it is 66.37\% and still at the same level as the baseline.
After prompt learning, the $\textit{Acc}_{\text{target}}$ of the prompted model also reaches 54.16\%, which is 0.06\% higher than the baseline.
In summary, using a shadow dataset with a similar distribution can achieve the adversarial goals in both attacks.

\begin{table}[!t]
\centering
\caption{
Influence of different datasets (STL-10).
Numbers in brackets are baselines.}
\label{table:stl10}
\setlength{\tabcolsep}{3pt}
\begin{tabular}{l c c c}
\toprule
Attack & \textit{ASR}/\textit{MR} & $\textit{Acc}_{\text{target}}$ & $\textit{Acc}_{\text{pre}}$ \\
\midrule
Targeted & 97.64 (17.02) & 54.16 (54.10) & 66.37 (69.76) \\
Untargeted & 74.74 (50.88) & 46.58 (54.10)& 66.11 (69.76) \\
\bottomrule
\end{tabular}
\end{table}

\revision{We further use UTKFace~\cite{ZSQ17}, a gender classification task (target label 1: female), to backdoor the model.
We conduct prompt learning on CIFAR-10 with the UTKFace backdoored model.
In this way, we aim to empirically observe how dramatically different shadow and downstream datasets may impact the attack performance.
Our attack achieves a 79.26\% $ASR$ but with 11.50\% of $Acc_{target}$.
Note that STL-10 and CIFAR-10 contain natural images with some objectives, while UTKFace only contains faces and focuses on gender classification.
We argue that our method is more transferable when the shadow dataset has similar categories and tasks, thereby preserving utility.
This is a practical assumption, as model providers typically have some insight into the downstream applications their clients are targeting.}

\begin{table}[t]
\centering
\caption{\method-attacked Instagram model on CIFAR-10.}
\label{table:attack_ins}
\setlength{\tabcolsep}{3pt}
\begin{tabular}{ l c  c  c}
\toprule
 & \textit{ASR} & $\textit{Acc}_{\text{target}}$ \\
\midrule
Baseline & 4.55 & 62.10 \\
\midrule
Targeted & 92.47 & 78.27 \\
\bottomrule
\end{tabular}
\end{table}

\mypara{Transfer to Different Models}
To show that our attack also applies to model architectures other than ResNet, we experimented on ResNeXt trained on 3.5B Instagram images (Instagram).
For Instagram, we follow the default settings in the paper and use CIFAR-10 as the target dataset.
Results in Table~\ref{table:attack_ins} show our attack's effectiveness, where \method achieves an $ASR$ of 92.47\% on the CIFAR-backdoored Instagram model.
Intuitively, our algorithm should also work for the vision-language model CLIP, specifically the image encoder.
However, as the pre-trained encoder is trained in a self-supervised way, the clean loss of the poisoning phase may be different.
Thus, the clean loss becomes the contrastive loss, where we add the text encoder and keep it fixed.
We leave this as future work.

\subsection{Ablation Study}
\label{section:ablation_study}

Here, we investigate factors including loss terms, number of iterations, quantity of poisoned samples, trigger size and placement, target label, and flipping strategies.
Both targeted and untargeted approaches are explored where applicable.
For all experiments in this section, we employ the ResNet18 model with the CIFAR-10 dataset.

\begin{figure}[!t]
\centering
\subfloat[\footnotesize{Targeted}]{\includegraphics[width=\columnwidth]{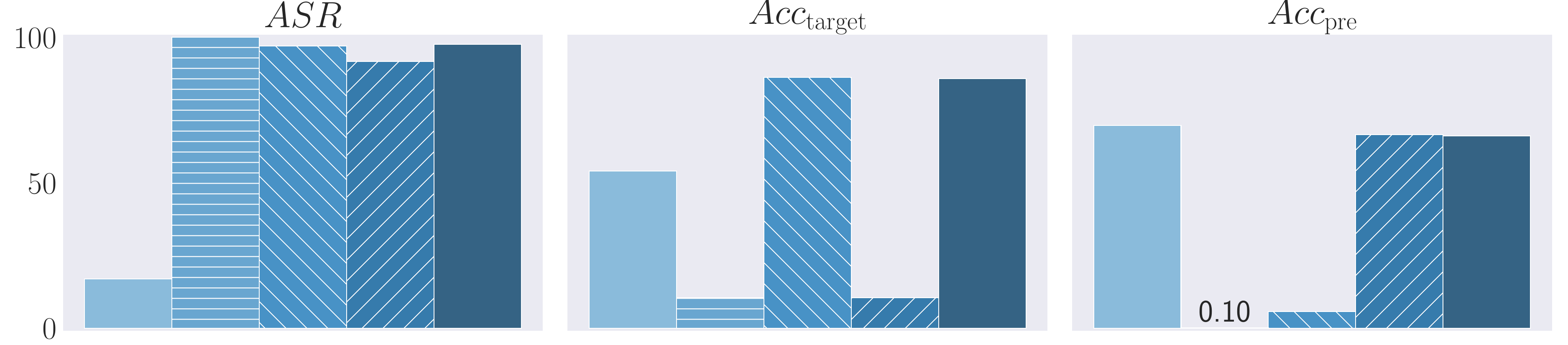}%
\label{figure:loss_target}}
\hfil
\subfloat[\footnotesize{Untargeted}]{\includegraphics[width=\columnwidth]{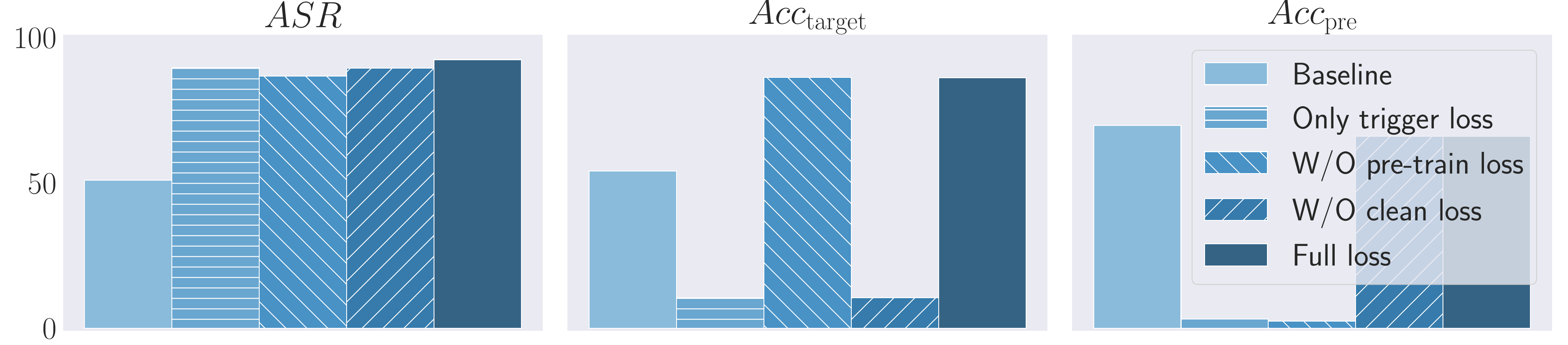}%
\label{figure:loss_untarget}}
\caption{Influence of the loss terms.
``Full loss'' represents the backdoored model with the poisoning loss ${\ell}_{\text{poisoning}}$ with $\alpha=1$, $\beta=1$, and $\gamma=1$ in Equation~\ref{equation:poisoning_loss} and the other three contain different loss terms in Equation~\ref{equation:poisoning_loss}.
``Only trigger loss'' means $\alpha=0$, $\beta=0$, and $\gamma=1$, ``W/O pre-train loss'' is $\alpha=0$, $\beta=1$, and $\gamma=1$ and ``W/O clean loss'' applies for $\alpha=1$, $\beta=0$, and $\gamma=1$.
}
\label{figure:loss}
\end{figure}

\mypara{Loss Terms}
We investigate the impact of different loss terms in Equation~\ref{equation:poisoning_loss}.
Results are shown in Figure~\ref{figure:loss}.
We observe that weighing more on the trigger loss ${\ell}_{D_t}$ can boost the attack's effectiveness.
For example, the \textit{ASR} of the targeted backdoored model with only the trigger loss reaches 100.00\%, achieving an 82.98\% performance gain over the baseline.
However, using only the trigger loss cannot satisfy our utility goal.
For instance, with only the trigger loss, the $\textit{Acc}_{\text{target}}$ and $\textit{Acc}_{\text{pre}}$ of the targeted backdoored model drop to 10.48\% and 3.13\%, while the baselines are 54.10\% and 69.76\%, respectively.
The results indicate that the pre-training loss ${\ell}_{D_p}$ and the clean loss ${\ell}_{D_c}$ are indispensable to maintain the utility of the backbone model on the pre-training task and the target task, respectively.
This finding is further supported by the observation that when the pre-training loss ${\ell}_{D_p}$ is excluded, the targeted backdoored model can achieve an 86.02\% $\textit{Acc}_{\text{target}}$ on the target task while only a 5.78\% $\textit{Acc}_{\text{pre}}$ on the pre-training task.
This highlights the importance of the pre-training loss in preserving the utility of the pre-training task.
In addition, if the clean loss ${\ell}_{D_c}$ is excluded, with the targeted backdoor attack, the $\textit{Acc}_{\text{pre}}$ increases to 66.54\% (the baseline is 69.76\%), while the $\textit{Acc}_{\text{target}}$ drops to 10.50\%, respectively.
Our results demonstrate the effectiveness of the trigger loss for the attack and the importance of the other two losses in contributing to the utility.
Further, Table~\ref{table:impact_of_lost_terms_appendix} shows additional results of different weight combinations among $\gamma$, $\beta$, and $\alpha$ for targeted attack, exhibiting similar conclusions.

\begin{table}[!t]
\centering
\caption{Influence of different loss weights.}
\label{table:impact_of_lost_terms_appendix}
\setlength{\tabcolsep}{3pt}
\begin{tabular}{lccc}
\toprule
Setting ($\gamma$:$\beta$:$\alpha$) & ASR & $Acc_{target}$ & $Acc_{clean}$ \\
\midrule
$D_c$ only (0:1:0) & 10.24 & 93.04 & 6.62 \\
$D_t$ only (1:0:0) & 100.00 & 10.48 & 0.10 \\
Weighted (1:0:5:0.5) & 98.82 & 87.10 & 65.67 \\
\bottomrule
\end{tabular}
\end{table}

\begin{figure}[!t]
\centering
\subfloat[\footnotesize{Targeted}]{\includegraphics[width=0.49\columnwidth]{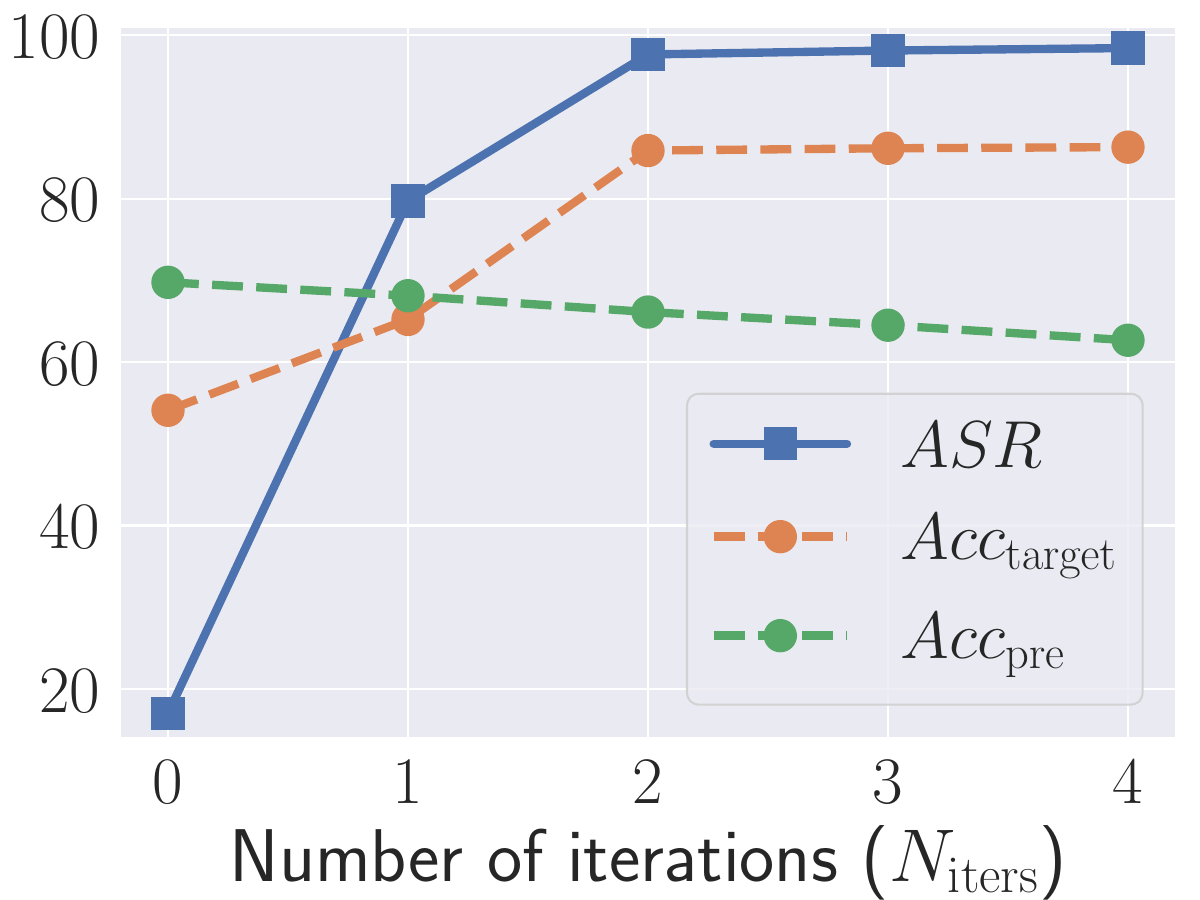}%
\label{figure:loops_target}}
\hfil
\subfloat[\footnotesize{Untargeted}]{\includegraphics[width=0.49\columnwidth]{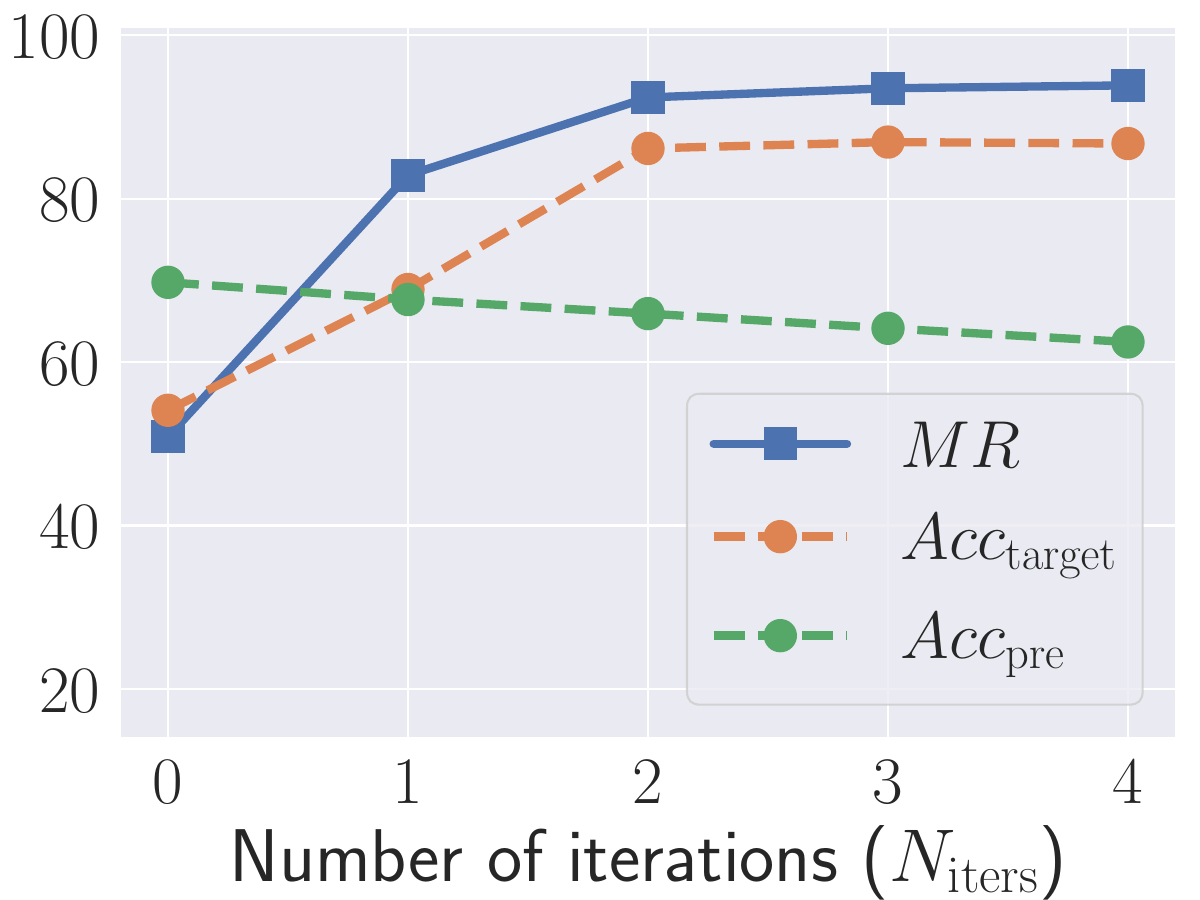}%
\label{figure:loops_untarget}}
\caption{Influence of the number of iterations.
}
\label{figure:loops}
\end{figure}

\begin{figure}[!t]
\centering
\subfloat[\footnotesize{Targeted}]{\includegraphics[width=0.49\columnwidth]{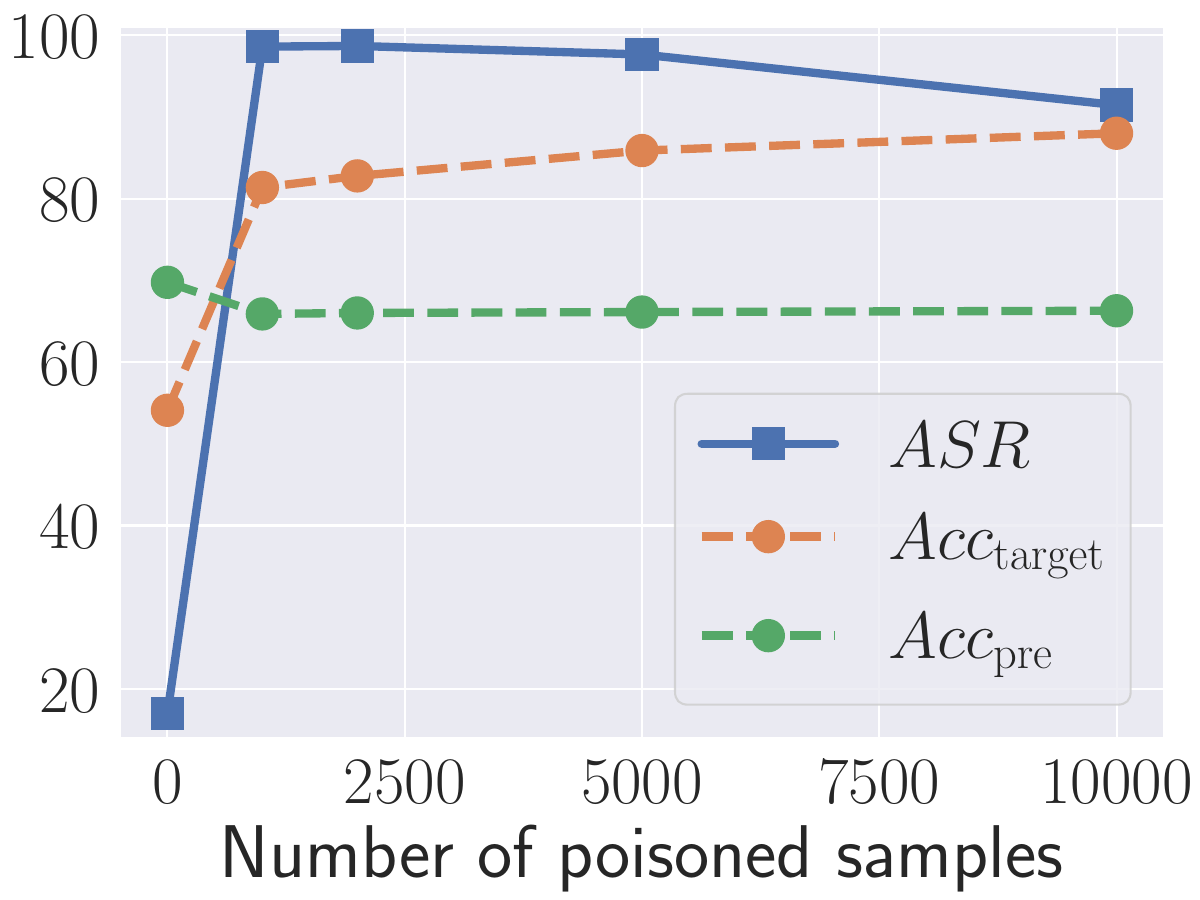}%
\label{figure:poisoned_samples_target}}
\hfil
\subfloat[\footnotesize{Untargeted}]{\includegraphics[width=0.49\columnwidth]{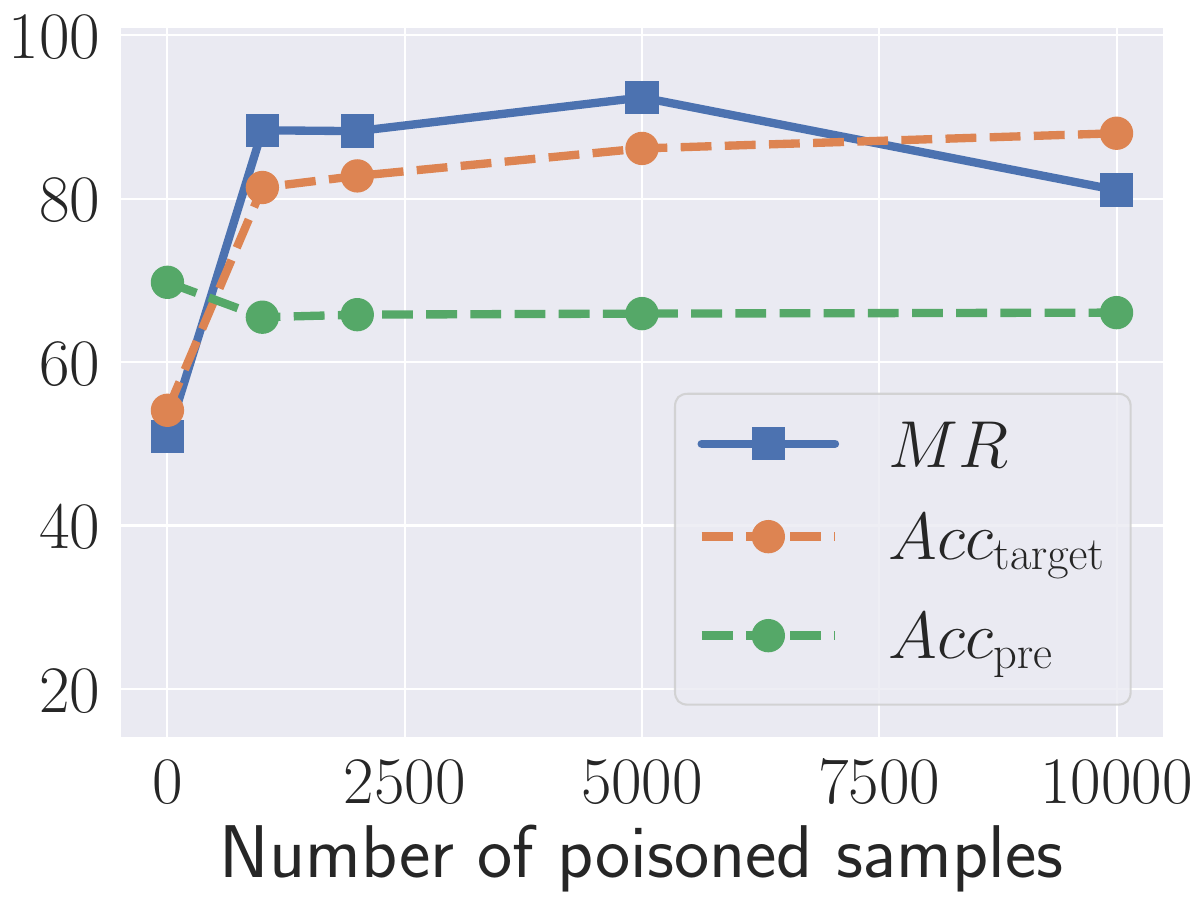}%
\label{figure:poisoned_samples_untarget}}
\caption{Influence of the number of poisoned samples.}
\label{figure:poisoned_samples}
\end{figure}

\mypara{Number of Iterations ($N_{\text{iters}}$)}
Figure~\ref{figure:loops} shows the performance of our two attacks with different numbers of iterations ($N_{\text{iters}}$ in Algorithm~\ref{algorithm:targeted}) ranging from 0 to 4.
When $N_{\text{iters}}$ is 0, it represents our baseline.
We find that the attacks perform well after only one iteration, which exemplifies the effectiveness of our attacks.
For example, the \textit{ASR} of the targeted backdoored model reaches 79.72\% in the first iteration, 62.70\% higher than the baseline.
With additional iterations, the attack performance first increases and then becomes stable in general.
We also observe that with more iterations, $\textit{Acc}_{\text{target}}$ increases, while $\textit{Acc}_{\text{pre}}$ slightly drops.
Therefore, we choose $N_{\text{iters}}=$2 to balance between utility and effectiveness for our attacks.

\mypara{Number of Poisoned Samples}
In our main experiments, we randomly sample 5,000 images from the downstream datasets to add triggers, i.e., the triggered dataset $D_t$, and construct the trigger loss, and another 5,000 images, i.e., the clean dataset $D_c$, to construct the clean loss.
To find out the influence of different numbers of poisoned samples on our two attacks, we randomly sample 1,000, 2,000, 5,000, and 10,000 images from CIFAR-10 and build the triggered dataset $D_t$.
We keep the number of clean samples to be the same as that of the poisoned ones to maintain the balance between $D_t$ and $D_c$.

Figure~\ref{figure:poisoned_samples} demonstrates the performance of our two attacks with these four different numbers of poisoned samples.
We find that our attacks can be characterized by an acceptable data resource requirement.
For example, with 1,000 poisoned samples, the \textit{ASR} of the targeted backdoored model can reach 98.60\%, 81.58\% higher than the baseline.
Given more poisoned samples, the attack performance improves.
But if there are too many poisoned samples, the attack performance may drop slightly.
For example, with 10,000 poisoned samples, the \textit{ASR} of the targeted backdoored model drops to 91.44\%, and the \textit{MR} of the untargeted backdoored model drops to 81.04\%, respectively.
The results indicate that injecting a proper number of poisoned samples is also important for the backdoor attack performance.

\begin{figure}[!t]
\centering
\subfloat[\footnotesize{Targeted}]{\includegraphics[width=\columnwidth]{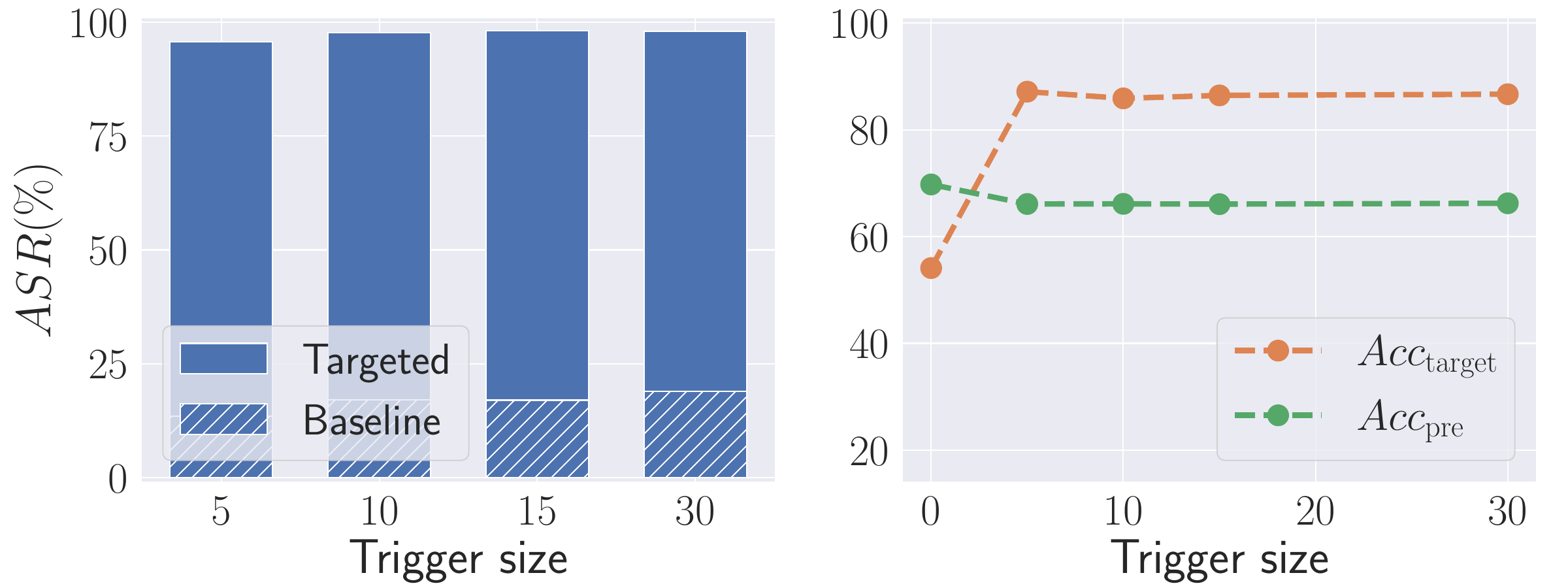}%
\label{figure:trigger_size_target}}
\hfil
\subfloat[\footnotesize{Untargeted}]{\includegraphics[width=\columnwidth]{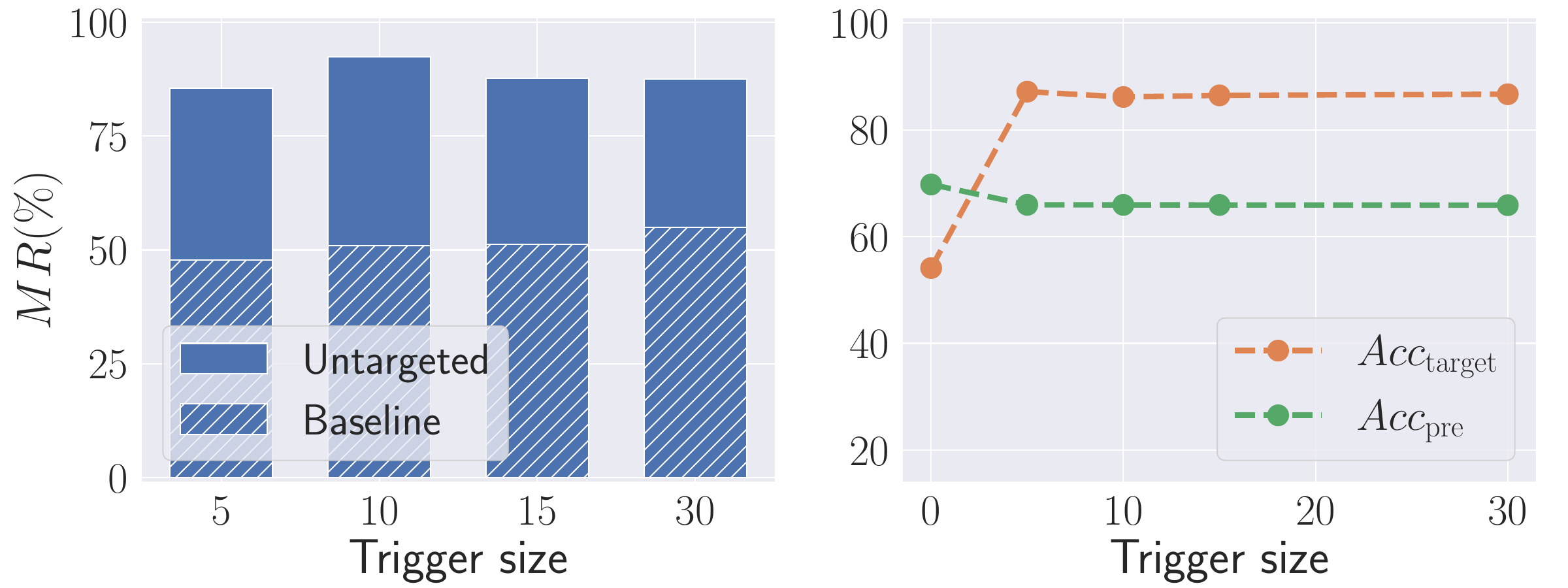}%
\label{figure:trigger_size_untarget}}
\caption{Influence of the trigger size.
For each attack, the left figure evaluates the attack performance, while the right illustrates the attack utility with different trigger sizes.
}
\label{figure:trigger_size}
\end{figure}

\mypara{Trigger Size}
We conduct both targeted and untargeted backdoor attacks with five different trigger sizes, i.e., 0, 5, 10, 15, and 30, to investigate the influence of the trigger size.
The size of 0 represents the clean backbone model, i.e., the baseline.
Note that all the triggers are at the same position, i.e., the bottom right corner.
We compute the \textit{ASR}/\textit{MR} baselines for each size, as the baseline performance varies with different trigger sizes.
For example, the baseline \textit{ASR} for the trigger size of 5 is 13.46\% while it is 17.02\% when the trigger size is 10.
Figure~\ref{figure:trigger_size} shows that the performance of both attacks improves with larger trigger sizes.
Take the targeted backdoor attack as an example; with larger trigger sizes, the \textit{ASR} of the backdoored model increases to 98\%, and the $\textit{Acc}_{\text{pre}}$ and $\textit{Acc}_{\text{target}}$ remain good.
This reveals that a larger trigger can usually lead to a better backdoor attack result.
However, if the trigger is too large, it is also easier to be detected.

\begin{table}[!t]
\centering
\caption{Influence of different trigger positions.}
\label{table:position}
\setlength{\tabcolsep}{3pt}
\begin{tabular}{l c c c c}
\toprule
Attack & Position & \textit{ASR}/\textit{MR} & $\textit{Acc}_{\text{target}}$ & $\textit{Acc}_{\text{pre}}$ \\
\midrule
\multirow{2}{*}{Targeted} & Center & 99.04 (10.74) & 86.68 & 66.17 \\
& Right bottom & 97.64 (17.02) & 87.09 & 71.96 \\
\midrule
\multirow{2}{*}{Untargeted} & Center & 88.60 (46.58) & 86.68 & 65.98 \\
& Right bottom & 92.38 (50.88) & 86.14 & 65.93  \\
\bottomrule
\end{tabular}
\end{table}

\mypara{Trigger Position}
We also investigate if trigger position may affect the attack performance.
To this end, we use the same trigger size, i.e., 10, and compare the performance of our two attacks with two different trigger positions, i.e., center and bottom right corner.
Table~\ref{table:position} shows the performance of \method with different trigger positions.
The numbers in the brackets are the baselines, i.e., the \textit{ASR/MR} of the clean model with a learned prompt on the triggered images with different trigger positions.
We observe that backdoored models with triggers at the center achieve better attack performance over the baselines.
For instance, with a trigger located at the center, the targeted backdoored model achieves a 99.04\% \textit{ASR}, achieving an 88.30\% gain over the baseline.
When the trigger is placed at the bottom right corner, the \textit{ASR} of the backdoored model is 80.62\%.
For the untargeted backdoor attack, the \textit{MR} with center triggers has a 42.02\% gain over the baseline, while with the trigger at the bottom right corner, the backdoored model achieves a comparable 41.50\% gain.
The results show that both attacks are robust to different trigger positions.

\begin{table}[!t]
\centering
\caption{Influence of different target labels.}
\label{table:target_label}
\setlength{\tabcolsep}{3pt}
\begin{tabular}{l c c c}
\toprule
Backbone Model & \textit{ASR} & $\textit{Acc}_{\text{target}}$ & $\textit{Acc}_{\text{pre}}$ \\
\midrule
ResNet18 & 96.90 & 86.56 & 66.10 \\
ResNet50 & 98.62 & 90.98 & 71.87 \\
BiT-M-RN50 & 98.88& 93.46& 72.04 \\
\bottomrule
\end{tabular}
\end{table}

\begin{table}[!t]
\centering
\caption{Influence of different label flipping strategies.}
\label{table:flip}
\setlength{\tabcolsep}{3pt}
\begin{tabular}{l c c c}
\toprule
Strategy & \textit{MR} & $\textit{Acc}_{\text{target}}$ & $\textit{Acc}_{\text{pre}}$ \\
\midrule
Random flipping & 65.06 & 81.76 & 65.88 \\
Next flipping & 92.38 & 86.14 & 65.93 \\
\bottomrule
\end{tabular}
\end{table}

\mypara{Target Label}
To show that different target labels do not affect the attack's effectiveness, we further evaluate our targeted attack targeting different labels.
Specifically, we target ``cat'' in CIFAR-10 and the detailed results are shown in Table~\ref{table:target_label}.
The results show that our attack performance is not affected by different target labels.
For example, the backdoored ResNet18 model on CIFAR-10 achieves 96.90\% $\textit{ASR}$ and maintains its utility when targeting ``cat'' instead of ``automobile.''
This further exemplifies the generalizability of our target attack.

\mypara{{Next Flipping} vs.\ {Random Flipping}}
To explore the influence of label-flipping strategies in untargeted backdoor attacks, we compare the attack performance between \emph{next flipping} and \emph{random flipping}.
Table~\ref{table:flip} shows the results.
We observe that both label-flipping strategies can achieve effectiveness and utility goals.
In particular, the next flipping generally leads to a higher attack performance than random flipping.
The \textit{MR} of the backdoored model by next flipping is 92.38\%, even 27.32\% higher than that by random flipping.
Meanwhile, both strategies maintain a good utility.
For instance, the $\textit{Acc}_{\text{target}}$ of the backdoored model with random flipping reaches 81.76\%, which is 27.66\% higher than the baseline.
This is similar to the case of the next flipping strategy.

\mypara{Summary}
In summary, our ablation study demonstrates that all three loss terms are critical for balancing effectiveness and utility.
While the attack is effective even with few poisoned samples and is robust to variations in trigger design and placement, a careful balance of optimization parameters is necessary to achieve the desired stealth and performance.

\section{Defense}
\label{section:defense}

We evaluate six state-of-the-art defenses against backdoor attacks.
Neural Cleanse~\cite{WYSLVZZ19}, ABS~\cite{LLTMAZ19}, and MNTD~\cite{XWLBGL21} for detecting backdoored models.
We treat the prompted model $F_{\theta, \phi}$ (see Section~\ref{section:preliminary}) as the defense target and check if the prompted model is backdoored.
NAD~\cite{LLKLLM212}, CLP~\cite{ZTLL22}, and D-BR~\cite{CWW22} aim to erase the backdoor trigger of a backdoored model.
Recall that the victim freely controls the visual prompt generation process; we target the backdoored backbone models $f_{\theta}$.
We measure our targeted and untargeted backdoored models on CIFAR-10 (six models in total) on these defenses.

\begin{table}[!t]
\centering
\caption{Neural Cleanse Anomaly Index of different backdoored models on CIFAR-10.
The threshold is 2.}
\label{table:nc}
\setlength{\tabcolsep}{3pt}
\begin{tabular}{l c c c}
\toprule
Attack & ResNet18 & ResNet50 & BiT-M-RN50 \\
\midrule
Targeted & 1.13 & 1.37 & 1.03 \\
Untargeted & 1.28 & 1.37 & 1.45 \\
\bottomrule
\end{tabular}
\end{table}

\begin{table}[!t]
\centering
\caption{MNTD classification accuracy of different backdoored models on CIFAR-10.}
\label{table:mntd}
\setlength{\tabcolsep}{3pt}
\begin{tabular}{l c c c}
\toprule
Attack & ResNet18 & ResNet50 & BiT-M-RN50 \\
\midrule
Targeted & 2\% & 2\% & 98\% \\
Untargeted & 6\% & 40\% & 100\% \\
\bottomrule
\end{tabular}
\end{table}

\begin{table*}[!t]
\centering
\caption{Performance of NAD-defensed backdoored backbone models on CIFAR-10.
Numbers in brackets show the attack performance before the defense.}
\label{table:nad_performance_and_utility}
\setlength{\tabcolsep}{3pt}
\begin{tabular}{l l c c c}
\toprule
Attack & Metrics & ResNet18 & ResNet50 & BiT-M-RN50 \\
\midrule
\multirow{3}{*}{Targeted} & \textit{ASR} & 96.58 (97.64) & 99.10 (98.66) & 96.00 (98.92) \\
& $\textit{Acc}_{\text{target}}$ & 76.70 (85.88) & 81.72 (91.04) & 88.28 (93.38) \\
& $\textit{Acc}_{\text{pre}}$ & 61.46 (66.12) & 65.40 (72.03) & 64.56 (72.11) \\
\midrule
\multirow{3}{*}{Untargeted} & \textit{MR} & 91.28 (92.38) & 86.72 (94.88) & 95.98 (97.74) \\
& $\textit{Acc}_{\text{target}}$ & 72.76 (86.14) & 85.06 (91.92) & 86.76 (94.02) \\
& $\textit{Acc}_{\text{pre}}$ & 59.31 (65.93) & 66.604 (71.81) & 64.72 (71.94) \\
\bottomrule
\end{tabular}
\end{table*}

\noindent \textbf{Neural Cleanse}~\cite{WYSLVZZ19} uses reverse engineering to detect whether a model is backdoored.
It generates triggers for each output label and predicts if any of them is a backdoor trigger by computing the Anomaly Index.
If the index is above the threshold of 2, the model is predicted to be backdoored.
We measure our targeted and untargeted backdoored models on CIFAR-10 (six models in total) via Neural Cleanse.
For each backdoored model, we reverse engineer a trigger for each label and compute the Anomaly Index following the default settings.
Results shown in Table~\ref{table:nc} indicate that all six models are predicted as clean models.
For instance, the Anomaly Index of the targeted backdoored BiT-M-RN50 model on CIFAR-10 is only 1.03, which is lower than the threshold.
We therefore conclude that Neural Cleanse cannot mitigate our attacks.

\noindent \textbf{ABS}~\cite{LLTMAZ19} is also based on reverse engineering.
It first analyzes the model's inner neuron behaviors, i.e., activation status, by introducing varied levels of stimulation based on a set of clean images.
Then it reverses the backdoor trigger to determine whether a neuron is truly compromised by leveraging the results of the stimulation analysis.
The authors of ABS acknowledge that the current version of ABS only works on CIFAR-10~\cite{LLTMAZ19}.
Thus, we use the supported set containing 50 CIFAR-10 seed images provided by the authors and run ABS against the six models, i.e., the targeted and untargeted backdoored models on CIFAR-10.
Results show that ABS predicts all six backdoored models to be clean.
We find that ABS cannot defend against our attacks as well.

\noindent \textbf{MNTD}~\cite{XWLBGL21} uses meta-learning to train a classifier to score the likelihood of a model being backdoored.
Following previous studies~\cite{JLG22, SWBMZ22}, we build 200 benign shadow models and 200 backdoored shadow models using the CIFAR-10 dataset.
The backdoored shadow models are trained with different triggers via jumbo learning.
Then, we train 50 sequential meta-classifiers based on the feature representations of these 400 shadow models and report their classification accuracy, representing the likelihood that the model is backdoored.
Following Salem et al.~\cite{SWBMZ22}, any model that scores higher than the threshold 0.0 is deemed backdoored.
For example, given an input model, if only 5 out of 50 meta-classifiers output scores higher than 0.0, the classification accuracy is 10\%, i.e., $\frac{5}{50} \times 100\%$.
Results in Table~\ref{table:mntd} show that the backdoored ResNet18 and ResNet50 models are less likely to be detected by MNTD, while the backdoored BiT-M-RN50 models have a high probability of being detected.

\begin{table}[!t]
\centering
\caption{Targeted Attack Performance and Utility of defensed ResNet18 Model.}
\label{table:defense_clp_and_dbr}
\setlength{\tabcolsep}{3pt}
\begin{tabular}{ l c  c  c}
\toprule
 & \textit{ASR} & $\textit{Acc}_{\text{target}}$ \\
\midrule
Targeted & 99.22 & 77.52 \\
\midrule
CLP Defensed & 13.52 & 52.10 \\
D-BR Defensed & 99.96 & 13.94 \\
\bottomrule
\end{tabular}
\end{table}

\noindent \textbf{NAD}~\cite{LLKLLM212} mitigates backdoor attacks by erasing backdoor triggers from a backdoored model via fine-tuning on clean data and attention distillation using a teacher-student framework.
As the victim controls the full process of prompt learning based on the frozen backbone model, it is natural to use NAD to defend against the backbone model $f_{\theta}$ instead of the prompted model $F_{\theta, \phi}$.
\revision{Specifically, we fine-tune the backdoored model $f_{\theta}$ on ImageNet-1k (50k) for 5 epochs with a learning rate of 0.001 and take it as the teacher model.
Then, we prune the backdoored model using attention distillation for 10 epochs with the same learning rate.}
We evaluate the attack performance and the utility based on the NAD-defensed backbone models.
The results are shown in Table~\ref{table:nad_performance_and_utility}.
The numbers in the bracket demonstrate the utility of the backdoored model before conducting NAD defense.
We observe that although the utility of the NAD-defensed models generally maintains at the same level as the backdoored model without defense, our attack still achieves a compatible attack success rate after defense.
For example, the $\textit{Acc}_{\text{target}}$ of the backdoored BiT-M-RN50 is 93.3\%, while that of the NAD-defensed model is 88.28\%.
The utility of the NAD-defensed model on the pre-training task slightly decreases by fine-tuning and pruning in the NAD defense process.
We consequently conclude that our attack is not successfully mitigated by NAD.

\noindent \textbf{CLP}~\cite{ZTLL22} and \textbf{D-BR}~\cite{CWW22} also aim to remove the backdoor from the model.
We target the targeted backdoored ResNet18 on the CIFAR10 dataset and then employ prompt learning on the defensed model to evaluate attack effectiveness and utility.
Results in Table~\ref{table:defense_clp_and_dbr} show that while CLP successfully mitigated the attack, it failed to maintain the utility of the target task.
Conversely, D-BR not only failed to remove the backdoor but also negatively affected the utility of the target task.
In conclusion, both defenses are ineffective at mitigating our attack while maintaining utility.

\mypara{Summary}
\revision{
We evaluate our approach against six state-of-the-art model-level defenses, including Neural Cleanse, ABS, MNTD, NAD, CLP, and D-BR.
Results demonstrate that most of our backdoored models can successfully bypass the defenses.
This is because most defenses rely on detecting abnormal behavior under trigger-only or global perturbations, whereas our co-activation mechanism keeps the backdoor dormant unless both the learned prompt and trigger are present.
That is, our \method is activated only when the learned prompt and the trigger are jointly present.
Possible defense directions could be prompt-agnostic behavioral consistency checks, decoupling-based tests for prompt-only/trigger-only activation, and cross-task anomaly analysis, which are left as future work.
}

\section{Related Work}
\label{section:related_work}

\mypara{Prompt Learning in Natural Language Processing}
Prompt learning, originating from natural language processing (NLP)~\cite{LYFJHN23}, involves creating task-specific templates to rephrase input text (e.g., ``\texttt{[INPUT]} Overall, I felt \texttt{[MASK]}.'') and mapping answers to desired labels using pre-trained language models (e.g., ``good/terrible'' for positive/negative).
Much effort has been taken in prompt engineering to design an effective textual prompt, i.e., discrete prompt, for the target downstream task~\cite{JXAN20, YNL21}.
However, hand-crafting the right prompt requires domain expertise and substantial trial-and-error effort.
Recent approaches~\cite{HKM21, LL21, LAC21, LZDDQYT21, HZDLS22, BOR22} have aimed to address this issue by learning a ``soft prompt,'' i.e., a continuous vector, through backpropagation while freezing the backbone model's parameters.
These include prompt tuning~\cite{LL21} and hybrid tuning~\cite{LZDDQYT21}.

\mypara{Prompt Learning in Computer Vision}
Visual prompt learning~\cite{EGS19, TMCEVH21, JTCCBHL22, BJSI22, CYCZL22, ZLZHL22, BGDGE22, YSBZ23, LSPC23} has emerged as a natural extension of prompt learning in NLP.
Visual prompt learning can be broadly categorized into two types.
The first category focuses on learning an input perturbation to shift the downstream data to the distribution used to train the pre-trained model~\cite{EGS19, BJSI22,LSPC23, HDCZWHY23}.
The second category aims to inject a small number of learnable parameters into the vision transformer~\cite{JTCCBHL22,BGDGE22,KKR23,XWCZLWZ23}.
However, these approaches require changes at the model level, which may raise security concerns among users.
In our paper, we thus focus on visual prompt learning from the first category.

\mypara{Backdoor Attacks and Their Defense}
Backdoor attack is a prominent security threat to machine learning models~\cite{CLLLS17, GDG17, LMALZWZ18, CT22, SWBMZ22, JLG22}.
It implants a trigger~\cite{ZPMJ21,SWBMZ22,LLWLHL21,SBZ20} into the target model during the training process, consequently causing the model to misclassify any input with the added trigger pattern to a specific target label.
Previous research has exemplified that backdoor attack can impact many real-world applications, including face recognition~\cite{CLLLS17}, image classification~\cite{YLZZ19,SSP20}, video processing~\cite{ZMZBCJ20}, NLP~\cite{CSBMSWZ21,SSTS20}, and, most recently, pre-trained encoders~\cite{JLG22,SJZLCSFYW21}.
Meanwhile, defense against backdoor attacks can be roughly divided into three detection levels, i.e., model-level~\cite{WYSLVZZ19, LLTMAZ19, XWLBGL21, CFZK19, HAS19, GWXDS19, LLKLLM212, ZTLL22, CWW22}, input-level~\cite{CJOK20, K21, GXWCRN19}, and dataset-level (i.e., if the training dataset is backdoored)~\cite{TLM18, HKSO21}.

\section{Conclusion}
\label{section:conclusion}

In this work, we explore the backdoor attack against an emerging machine learning paradigm, namely prompt learning.
Existing backdoor attacks against prompt learning focus on injecting the backdoor into the prompt~\cite{DZLLW22, HZBSZ232}.
We are the first to propose \method, the targeted/untargeted backdoor attack against the backbone model in visual prompt learning.
We develop a bi-level optimization-based approach to inject a backdoor into the backbone model, where the backdoor can be activated when prompt learning is performed for the target task.
We evaluate our \method using three benchmark datasets on three different model frameworks, exhibiting its effectiveness, stealthiness, and transferability.
Experiments show that both attacks can achieve their adversarial goal while preserving the model's utility.
We further show that most of our backdoored models can bypass six state-of-the-art backdoor defense mechanisms, requesting new defenses for our backdoor attacks.

\section*{Limitations}

Our attack conditions on two factors: knowledge of the downstream task and adherence to the label mapping scheme by the victim.
Regarding the first condition, it is typical in transfer learning, such as visual prompt learning, for the model provider to collaborate with the victim to select the most suitable model.
Consequently, attackers may gain insight into the downstream task.
Experiments in Section~\ref{section:transfer} further show that our attack can be generalized to out-of-distribution datasets of a downstream task.
In this sense, our attack also supports scenarios when the attacker cannot obtain the training data for the downstream task.
Concerning the second condition, we acknowledge that our attack may not yield optimal outcomes if the victim does not adhere to the predefined label mapping scheme.
However, we emphasize that our attack targets victims lacking advanced machine learning expertise, as outlined in the threat model (Section~\ref{section:threat_model}).
In such cases, it is plausible that they will follow the mapping scheme recommended by the model provider rather than experimenting with alternative schemes.

\section*{Impact Statement}

Prompt learning has emerged as a trending topic in machine learning, gaining significant attention due to its time and computational efficiency, particularly in industrial applications.
However, this paradigm introduces critical security risks that must be addressed.
Our work highlights a potential threat: the backdooring of backbone models within the prompt learning framework.
Given that our attack method is both simple and highly effective, it poses a significant danger if exploited by malicious actors.
By exposing this vulnerability, our research contributes to the development of more robust defenses in real-world scenarios.
\revision{Furthermore, we aim to advance the attack-defense cycle by releasing our code publicly for reproducibility, ultimately facilitating the development of effective defenses.
The code is released under the MIT license.
In addition, the repository README includes a responsible-use statement clarifying that the release is intended for reproducibility and defensive research, and that users are expected not to use the code to develop, deploy, distribute, or facilitate malicious models, attacks, unauthorized access, evasion, or other harmful activity.}

\section*{Acknowledgment}

This work has been partially supported by the National Key Research and Development Program of China (No. 2025YFB3110200), Japan society for the promotion of science (JSPS), Grants-in-Aid for Scientific Research 26K21215, the Guangdong Basic and Applied Basic Research Foundation (No. 2026A1515030046), and the State Key Laboratory of Internet Architecture, Tsinghua University (No. HLW2025ZD14).

\begin{small}
\bibliographystyle{plain}
\bibliography{normal_generated_py3}

@inproceedings{SRS17,
author = {Congzheng Song and Thomas Ristenpart and Vitaly Shmatikov},
title = {{Machine Learning Models that Remember Too Much}},
booktitle = {{ACM SIGSAC Conference on Computer and Communications Security (CCS)}},
pages = {587-601},
publisher = {ACM},
year = {2017}
}

@inproceedings{LMALZWZ18,
author = {Yingqi Liu and Shiqing Ma and Yousra Aafer and Wen-Chuan Lee and Juan Zhai and Weihang Wang and Xiangyu Zhang},
title = {{Trojaning Attack on Neural Networks}},
booktitle = {{Network and Distributed System Security Symposium (NDSS)}},
publisher = {Internet Society},
year = {2018}
}

@inproceedings{YLZZ19,
author = {Yuanshun Yao and Huiying Li and Haitao Zheng and Ben Y. Zhao},
title = {{Latent Backdoor Attacks on Deep Neural Networks}},
booktitle = {{ACM SIGSAC Conference on Computer and Communications Security (CCS)}},
pages = {2041-2055},
publisher = {ACM},
year = {2019}
}

@inproceedings{WYSLVZZ19,
author = {Bolun Wang and Yuanshun Yao and Shawn Shan and Huiying Li and Bimal Viswanath and Haitao Zheng and Ben Y. Zhao},
title = {{Neural Cleanse: Identifying and Mitigating Backdoor Attacks in Neural Networks}},
booktitle = {{IEEE Symposium on Security and Privacy (S\&P)}},
pages = {707-723},
publisher = {IEEE},
year = {2019}
}

@inproceedings{LLTMAZ19,
author = {Yingqi Liu and Wen-Chuan Lee and Guanhong Tao and Shiqing Ma and Yousra Aafer and Xiangyu Zhang},
title = {{ABS: Scanning Neural Networks for Back-Doors by Artificial Brain Stimulation}},
booktitle = {{ACM SIGSAC Conference on Computer and Communications Security (CCS)}},
pages = {1265-1282},
publisher = {ACM},
year = {2019}
}

@inproceedings{GXWCRN19,
author = {Yansong Gao and Change Xu and Derui Wang and Shiping Chen and Damith C Ranasinghe and Surya Nepal},
title = {{STRIP: A Defence Against Trojan Attacks on Deep Neural Networks}},
booktitle = {{Annual Computer Security Applications Conference (ACSAC)}},
pages = {113-125},
publisher = {ACM},
year = {2019}
}

@inproceedings{HZRS16,
author = {Kaiming He and Xiangyu Zhang and Shaoqing Ren and Jian Sun},
title = {{Deep Residual Learning for Image Recognition}},
booktitle = {{IEEE Conference on Computer Vision and Pattern Recognition (CVPR)}},
pages = {770-778},
publisher = {IEEE},
year = {2016}
}

@inproceedings{XWLBGL21,
author = {Xiaojun Xu and Qi Wang and Huichen Li and Nikita Borisov and Carl A. Gunter and Bo Li},
title = {{Detecting AI Trojans Using Meta Neural Analysis}},
booktitle = {{IEEE Symposium on Security and Privacy (S\&P)}},
publisher = {IEEE},
year = {2021}
}

@inproceedings{SSP20,
author = {Aniruddha Saha and Akshayvarun Subramanya and Hamed Pirsiavash},
title = {{Hidden Trigger Backdoor Attacks}},
booktitle = {{AAAI Conference on Artificial Intelligence (AAAI)}},
pages = {11957-11965},
publisher = {AAAI},
year = {2020}
}

@inproceedings{ZMZBCJ20,
author = {Shihao Zhao and Xingjun Ma and Xiang Zheng and James Bailey and Jingjing Chen and Yu-Gang Jiang},
title = {{Clean-Label Backdoor Attacks on Video Recognition Models}},
booktitle = {{IEEE Conference on Computer Vision and Pattern Recognition (CVPR)}},
pages = {14443-144528},
publisher = {IEEE},
year = {2020}
}

@inproceedings{ZSQ17,
author = {Zhifei Zhang and Yang Song and Hairong Qi},
title = {{Age Progression/Regression by Conditional Adversarial Autoencoder}},
booktitle = {{IEEE Conference on Computer Vision and Pattern Recognition (CVPR)}},
pages = {4352-4360},
publisher = {IEEE},
year = {2017}
}

@inproceedings{CNL11,
author = {Adam Coates and Andrew Y. Ng and Honglak Lee},
title = {{An Analysis of Single-Layer Networks in Unsupervised Feature Learning}},
booktitle = {{International Conference on Artificial Intelligence and Statistics (AISTATS)}},
pages = {215-223},
publisher = {JMLR},
year = {2011}
}

@inproceedings{EGS19,
author = {Gamaleldin F. Elsayed and Ian J. Goodfellow and Jascha Sohl{-}Dickstein},
title = {{Adversarial Reprogramming of Neural Networks}},
booktitle = {{International Conference on Learning Representations (ICLR)}},
year = {2019}
}

@inproceedings{KBZPYGH20,
author = {Alexander Kolesnikov and Lucas Beyer and Xiaohua Zhai and Joan Puigcerver and Jessica Yung and Sylvain Gelly and Neil Houlsby},
title = {{Big Transfer (BiT): General Visual Representation Learning}},
booktitle = {{European Conference on Computer Vision (ECCV)}},
pages = {491-507},
publisher = {Springer},
year = {2020}
}

@inproceedings{JLG22,
author = {Jinyuan Jia and Yupei Liu and Neil Zhenqiang Gong},
title = {{BadEncoder: Backdoor Attacks to Pre-trained Encoders in Self-Supervised Learning}},
booktitle = {{IEEE Symposium on Security and Privacy (S\&P)}},
publisher = {IEEE},
year = {2022}
}

@inproceedings{CSBMSWZ21,
author = {Xiaoyi Chen and Ahmed Salem and Michael Backes and Shiqing Ma and Qingni Shen and Zhonghai Wu and Yang Zhang},
title = {{BadNL: Backdoor Attacks Against NLP Models with Semantic-preserving Improvements}},
booktitle = {{Annual Computer Security Applications Conference (ACSAC)}},
pages = {554-569},
publisher = {ACSAC},
year = {2021}
}

@inproceedings{DDSLLF09,
author = {Jia Deng and Wei Dong and Richard Socher and Li{-}Jia Li and Kai Li and Li Fei{-}Fei},
title = {{ImageNet: {A} large-scale hierarchical image database}},
booktitle = {{IEEE Conference on Computer Vision and Pattern Recognition (CVPR)}},
pages = {248-255},
publisher = {IEEE},
year = {2009}
}

@inproceedings{RKHRGASAMCKS21,
author = {Alec Radford and Jong Wook Kim and Chris Hallacy and Aditya Ramesh and Gabriel Goh and Sandhini Agarwal and Girish Sastry and Amanda Askell and Pamela Mishkin and Jack Clark and Gretchen Krueger and Ilya Sutskever},
title = {{Learning Transferable Visual Models From Natural Language Supervision}},
booktitle = {{International Conference on Machine Learning (ICML)}},
pages = {8748-8763},
publisher = {PMLR},
year = {2021}
}

@inproceedings{DBKWZUDMHGUH21,
author = {Alexey Dosovitskiy and Lucas Beyer and Alexander Kolesnikov and Dirk Weissenborn and Xiaohua Zhai and Thomas Unterthiner and Mostafa Dehghani and Matthias Minderer and Georg Heigold and Sylvain Gelly and Jakob Uszkoreit and Neil Houlsby},
title = {{An Image is Worth 16x16 Words: Transformers for Image Recognition at Scale}},
booktitle = {{International Conference on Learning Representations (ICLR)}},
year = {2021}
}

@inproceedings{SWBMZ22,
author = {Ahmed Salem and Rui Wen and Michael Backes and Shiqing Ma and Yang Zhang},
title = {{Dynamic Backdoor Attacks Against Machine Learning Models}},
booktitle = {{IEEE European Symposium on Security and Privacy (Euro S\&P)}},
pages = {703-718},
publisher = {IEEE},
year = {2022}
}

@inproceedings{NWCBWN11,
author = {Yuval Netzer and Tao Wang and Adam Coates and Alessandro Bissacco and Bo Wu and Andrew Y. Ng},
title = {{Reading Digits in Natural Images with Unsupervised Feature Learning}},
booktitle = {{Annual Conference on Neural Information Processing Systems (NIPS)}},
publisher = {NIPS},
year = {2011}
}

@inproceedings{TLM18,
author = {Brandon Tran and Jerry Li and Aleksander Madry},
title = {{Spectral Signatures in Backdoor Attacks}},
booktitle = {{Annual Conference on Neural Information Processing Systems (NeurIPS)}},
pages = {8011-8021},
publisher = {NeurIPS},
year = {2018}
}

@inproceedings{SJZLCSFYW21,
author = {Lujia Shen and Shouling Ji and Xuhong Zhang and Jinfeng Li and Jing Chen and Jie Shi and Chengfang Fang and Jianwei Yin and Ting Wang},
title = {{Backdoor Pre-trained Models Can Transfer to All}},
booktitle = {{ACM SIGSAC Conference on Computer and Communications Security (CCS)}},
pages = {3141-3158},
publisher = {ACM},
year = {2021}
}

@inproceedings{LL21,
author = {Xiang Lisa Li and Percy Liang},
title = {{Prefix-Tuning: Optimizing Continuous Prompts for Generation}},
booktitle = {{Annual Meeting of the Association for Computational Linguistics and International Joint Conference on Natural Language Processing (ACL/IJCNLP)}},
pages = {4582-4597},
publisher = {ACL},
year = {2021}
}

@inproceedings{LAC21,
author = {Brian Lester and Rami Al{-}Rfou and Noah Constant},
title = {{The Power of Scale for Parameter-Efficient Prompt Tuning}},
booktitle = {{Conference on Empirical Methods in Natural Language Processing (EMNLP)}},
pages = {3045-3059},
publisher = {ACL},
year = {2021}
}

@inproceedings{TMCEVH21,
author = {Maria Tsimpoukelli and Jacob Menick and Serkan Cabi and S. M. Ali Eslami and Oriol Vinyals and Felix Hill},
title = {{Multimodal Few-Shot Learning with Frozen Language Models}},
booktitle = {{Annual Conference on Neural Information Processing Systems (NeurIPS)}},
pages = {200-212},
publisher = {NeurIPS},
year = {2021}
}

@inproceedings{QE21,
author = {Guanghui Qin and Jason Eisner},
title = {{Learning How to Ask: Querying LMs with Mixtures of Soft Prompts}},
booktitle = {{Conference of the North American Chapter of the Association for Computational Linguistics: Human Language Technologies (NAACL-HLT)}},
pages = {5203-5212},
publisher = {ACL},
year = {2021}
}

@inproceedings{CFZK19,
author = {Huili Chen and Cheng Fu and Jishen Zhao and Farinaz Koushanfar},
title = {{DeepInspect: {A} Black-box Trojan Detection and Mitigation Framework for Deep Neural Networks}},
booktitle = {{International Joint Conferences on Artifical Intelligence (IJCAI)}},
pages = {4658-4664},
publisher = {IJCAI},
year = {2019}
}

@inproceedings{ZPMJ21,
author = {Yi Zeng and Won Park and Z. Morley Mao and Ruoxi Jia},
title = {{Rethinking the Backdoor Attacks' Triggers: {A} Frequency Perspective}},
booktitle = {{IEEE International Conference on Computer Vision (ICCV)}},
pages = {16453-16461},
publisher = {IEEE},
year = {2021}
}

@inproceedings{LLWLHL21,
author = {Yuezun Li and Yiming Li and Baoyuan Wu and Longkang Li and Ran He and Siwei Lyu},
title = {{Invisible Backdoor Attack with Sample-Specific Triggers}},
booktitle = {{IEEE International Conference on Computer Vision (ICCV)}},
pages = {16443-16452},
publisher = {IEEE},
year = {2021}
}

@inproceedings{LLKLLM212,
author = {Yige Li and Xixiang Lyu and Nodens Koren and Lingjuan Lyu and Bo Li and Xingjun Ma},
title = {{Neural Attention Distillation: Erasing Backdoor Triggers from Deep Neural Networks}},
booktitle = {{International Conference on Learning Representations (ICLR)}},
year = {2021}
}

@inproceedings{LCYLRBS20,
author = {Hao Li and Pratik Chaudhari and Hao Yang and Michael Lam and Avinash Ravichandran and Rahul Bhotika and Stefano Soatto},
title = {{Rethinking the Hyperparameters for Fine-tuning}},
booktitle = {{International Conference on Learning Representations (ICLR)}},
year = {2020}
}

@inproceedings{DZLLW22,
author = {Wei Du and Yichun Zhao and Boqun Li and Gongshen Liu and Shilin Wang},
title = {{PPT: Backdoor Attacks on Pre-trained Models via Poisoned Prompt Tuning}},
booktitle = {{International Joint Conferences on Artifical Intelligence (IJCAI)}},
pages = {680-686},
publisher = {IJCAI},
year = {2022}
}

@inproceedings{JTCCBHL22,
author = {Menglin Jia and Luming Tang and Bor{-}Chun Chen and Claire Cardie and Serge J. Belongie and Bharath Hariharan and Ser{-}Nam Lim},
title = {{Visual Prompt Tuning}},
booktitle = {{European Conference on Computer Vision (ECCV)}},
pages = {709-727},
publisher = {Springer},
year = {2022}
}

@inproceedings{CJOK20,
author = {Seung Ju Cho and Tae Joon Jun and Byungsoo Oh and Daeyoung Kim},
title = {{{DAPAS} : Denoising Autoencoder to Prevent Adversarial attack in Semantic Segmentation}},
booktitle = {{International Joint Conference on Neural Networks (IJCNN)}},
pages = {1-8},
publisher = {IEEE},
year = {2020}
}

@inproceedings{YNL21,
author = {Weizhe Yuan and Graham Neubig and Pengfei Liu},
title = {{BARTScore: Evaluating Generated Text as Text Generation}},
booktitle = {{Annual Conference on Neural Information Processing Systems (NeurIPS)}},
pages = {27263-27277},
publisher = {NeurIPS},
year = {2021}
}

@inproceedings{HKSO21,
author = {Jonathan Hayase and Weihao Kong and Raghav Somani and Sewoong Oh},
title = {{SPECTRE: Defending Against Backdoor Attacks Using Robust Statistics}},
booktitle = {{International Conference on Machine Learning (ICML)}},
publisher = {JMLR},
year = {2021}
}

@inproceedings{LH17,
author = {Ilya Loshchilov and Frank Hutter},
title = {{SGDR: Stochastic Gradient Descent with Warm Restarts}},
booktitle = {{International Conference on Learning Representations (ICLR)}},
year = {2017}
}

@inproceedings{LSPC23,
author = {Weihuang Liu and Xi Shen and Chi{-}Man Pun and Xiaodong Cun},
title = {{Explicit Visual Prompting for Low-Level Structure Segmentations}},
booktitle = {{IEEE Conference on Computer Vision and Pattern Recognition (CVPR)}},
publisher = {IEEE},
year = {2023}
}

@inproceedings{HDCZWHY23,
author = {Qidong Huang and Xiaoyi Dong and Dongdong Chen and Weiming Zhang and Feifei Wang and Gang Hua and Nenghai Yu},
title = {{Diversity-Aware Meta Visual Prompting}},
booktitle = {{IEEE Conference on Computer Vision and Pattern Recognition (CVPR)}},
publisher = {IEEE},
year = {2023}
}

@inproceedings{BGDGE22,
author = {Amir Bar and Yossi Gandelsman and Trevor Darrell and Amir Globerson and Alexei A. Efros},
title = {{Visual Prompting via Image Inpainting}},
booktitle = {{Annual Conference on Neural Information Processing Systems (NeurIPS)}},
pages = {25005-25017},
publisher = {NeurIPS},
year = {2022}
}

@inproceedings{HKM21,
author = {Karen Hambardzumyan and Hrant Khachatrian and Jonathan May},
title = {{{WARP:} Word-level Adversarial ReProgramming}},
booktitle = {{Annual Meeting of the Association for Computational Linguistics and International Joint Conference on Natural Language Processing (ACL/IJCNLP)}},
pages = {4921-4933},
publisher = {ACL},
year = {2021}
}

@inproceedings{STKP22,
author = {Aniruddha Saha and Ajinkya Tejankar and Soroush Abbasi Koohpayegani and Hamed Pirsiavash},
title = {{Backdoor Attacks on Self-Supervised Learning}},
booktitle = {{IEEE Conference on Computer Vision and Pattern Recognition (CVPR)}},
pages = {13327-13336},
publisher = {IEEE},
year = {2022}
}

@inproceedings{CT22,
author = {Nicholas Carlini and Andreas Terzis},
title = {{Poisoning and Backdooring Contrastive Learning}},
booktitle = {{International Conference on Learning Representations (ICLR)}},
year = {2022}
}

@inproceedings{CWW22,
author = {Weixin Chen and Baoyuan Wu and Haoqian Wang},
title = {{Effective Backdoor Defense by Exploiting Sensitivity of Poisoned Samples}},
booktitle = {{Annual Conference on Neural Information Processing Systems (NeurIPS)}},
publisher = {NeurIPS},
year = {2022}
}

@inproceedings{ZTLL22,
author = {Runkai Zheng and Rongjun Tang and Jianze Li and Li Liu},
title = {{Data-Free Backdoor Removal Based on Channel Lipschitzness}},
booktitle = {{European Conference on Computer Vision (ECCV)}},
pages = {175-191},
publisher = {Springer},
year = {2022}
}

@article{JXAN20,
author = {Zhengbao Jiang and Frank F. Xu and Jun Araki and Graham Neubig},
title = {{How Can We Know What Language Models Know}},
journal = {{Transactions of the Association for Computational Linguistics}},
publisher = {The MIT Press},
year = {2020}
}

@article{HBDB19,
author = {Patrick Helber and Benjamin Bischke and Andreas Dengel and Damian Borth},
title = {{Eurosat: A Novel Dataset and Deep Learning Benchmark for Land Use and Land Cover Classification}},
journal = {{IEEE Journal of Selected Topics in Applied Earth Observations and Remote Sensing}},
publisher = {IEEE},
year = {2019}
}

@article{HZDLS22,
author = {Xu Han and Weilin Zhao and Ning Ding and Zhiyuan Liu and Maosong Sun},
title = {{PTR: Prompt Tuning with Rules for Text Classification}},
journal = {{{AI} Open}},
publisher = {Elsevier},
year = {2022}
}

@article{BOR22,
author = {Eyal Ben{-}David and Nadav Oved and Roi Reichart},
title = {{PADA: Example-based Prompt Learning for on-the-fly Adaptation to Unseen Domains}},
journal = {{Transactions of the Association for Computational Linguistics}},
publisher = {The MIT Press},
year = {2022}
}

@article{K21,
author = {Hyun Kwon},
title = {{Defending Deep Neural Networks against Backdoor Attack by Using De-trigger Autoencoder}},
journal = {{IEEE Access}},
publisher = {IEEE},
year = {2021}
}

@article{LZLFY22,
author = {Guangrui Liu and Weizhe Zhang and Xinjie Li and Kaisheng Fan and Shui Yu},
title = {{VulnerGAN: A Backdoor Attack Through Vulnerability Amplification Against Machine Learning-based Network Intrusion Detection Systems}},
journal = {{Science China Information Sciences}},
publisher = {Springer},
year = {2022}
}

@article{LYFJHN23,
author = {Pengfei Liu and Weizhe Yuan and Jinlan Fu and Zhengbao Jiang and Hiroaki Hayashi and Graham Neubig},
title = {{Pre-train, Prompt, and Predict: {A} Systematic Survey of Prompting Methods in Natural Language Processing}},
journal = {{ACM Computing Surveys}},
publisher = {ACM},
year = {2023}
}

@article{PY09,
author = {Sinno Jialin Pan and Qiang Yang},
title = {{A Survey on Transfer Learning}},
journal = {{IEEE Transactions on Knowledge and Data Engineering}},
publisher = {IEEE},
year = {2009}
}

@article{GDG17,
author = {Tianyu Gu and Brendan Dolan-Gavitt and Siddharth Grag},
title = {{Badnets: Identifying Vulnerabilities in the Machine Learning Model Supply Chain}},
journal = {{CoRR abs/1708.06733}},
year = {2017}
}

@article{SSTS20,
author = {Roei Schuster and Congzheng Song and Eran Tromer and Vitaly Shmatikov},
title = {{You Autocomplete Me: Poisoning Vulnerabilities in Neural Code Completion}},
journal = {{CoRR abs/2007.02220}},
year = {2020}
}

@article{SBZ20,
author = {Ahmed Salem and Michael Backes and Yang Zhang},
title = {{Don't Trigger Me! A Triggerless Backdoor Attack Against Deep Neural Networks}},
journal = {{CoRR abs/2010.03282}},
year = {2020}
}

@article{PWSCL21,
author = {Or Patashnik and Zongze Wu and Eli Shechtman and Daniel Cohen{-}Or and Dani Lischinski},
title = {{StyleCLIP: Text-Driven Manipulation of StyleGAN Imagery}},
journal = {{CoRR abs/2103.17249}},
year = {2021}
}

@article{KTN20,
author = {Virapat Kieuvongngam and Bowen Tan and Yiming Niu},
title = {{Automatic Text Summarization of {COVID-19} Medical Research Articles using {BERT} and {GPT-2}}},
journal = {{CoRR abs/2006.01997}},
year = {2020}
}

@article{MHB21,
author = {Ron Mokady and Amir Hertz and Amit H. Bermano},
title = {{ClipCap: {CLIP} Prefix for Image Captioning}},
journal = {{CoRR abs/2111.09734}},
year = {2021}
}

@article{DISFHS20,
author = {Jesse Dodge and Gabriel Ilharco and Roy Schwartz and Ali Farhadi and Hannaneh Hajishirzi and Noah A. Smith},
title = {{Fine-Tuning Pretrained Language Models: Weight Initializations, Data Orders, and Early Stopping}},
journal = {{CoRR abs/2002.06305}},
year = {2020}
}

@article{CLLLS17,
author = {Xinyun Chen and Chang Liu and Bo Li and Kimberly Lu and Dawn Song},
title = {{Targeted Backdoor Attacks on Deep Learning Systems Using Data Poisoning}},
journal = {{CoRR abs/1712.05526}},
year = {2017}
}

@article{GWXDS19,
author = {Wenbo Guo and Lun Wang and Xinyu Xing and Min Du and Dawn Song},
title = {{{TABOR:} {A} Highly Accurate Approach to Inspecting and Restoring Trojan Backdoors in {AI} Systems}},
journal = {{CoRR abs/1908.01763}},
year = {2019}
}

@article{HAS19,
author = {Xijie Huang and Moustafa Alzantot and Mani B. Srivastava},
title = {{NeuronInspect: Detecting Backdoors in Neural Networks via Output Explanations}},
journal = {{CoRR abs/1911.07399}},
year = {2019}
}

@article{LZDDQYT21,
author = {Xiao Liu and Yanan Zheng and Zhengxiao Du and Ming Ding and Yujie Qian and Zhilin Yang and Jie Tang},
title = {{GPT Understands, Too}},
journal = {{CoRR abs/2103.10385}},
year = {2021}
}

@article{BJSI22,
author = {Hyojin Bahng and Ali Jahanian and Swami Sankaranarayanan and Phillip Isola},
title = {{Exploring Visual Prompts for Adapting Large-scale Models}},
journal = {{CoRR abs/2203.17274}},
year = {2022}
}

@article{CYCZL22,
author = {Aochuan Chen and Yuguang Yao and Pin{-}Yu Chen and Yihua Zhang and Sijia Liu},
title = {{Understanding and Improving Visual Prompting: A Label-Mapping Perspective}},
journal = {{CoRR abs/2211.11635}},
year = {2022}
}

@article{YSBZ23,
author = {Ziqing Yang and Zeyang Sha and Michael Backes and Yang Zhang},
title = {{From Visual Prompt Learning to Zero-Shot Transfer: Mapping Is All You Need}},
journal = {{CoRR abs/2303.05266}},
year = {2023}
}

@article{KKR23,
author = {Minsu Kim and Hyung{-}Il Kim and Yong Man Ro},
title = {{Prompt Tuning of Deep Neural Networks for Speaker-adaptive Visual Speech Recognition}},
journal = {{CoRR abs/2302.08102}},
year = {2023}
}

@article{XWCZLWZ23,
author = {Yinghui Xing and Qirui Wu and De Cheng and Shizhou Zhang and Guoqiang Liang and Peng Wang and Yanning Zhang},
title = {{Dual Modality Prompt Tuning for Vision-Language Pre-Trained Model}},
journal = {{CoRR abs/2208.08340}},
year = {2023}
}

@article{ZLZHL22,
author = {Yuhang Zang and Wei Li and Kaiyang Zhou and Chen Huang and Chen Change Loy},
title = {{Unified Vision and Language Prompt Learning}},
journal = {{CoRR abs/2210.07225}},
year = {2022}
}

@article{HZBSZ232,
author = {Hai Huang and Zhengyu Zhao and Michael Backes and Yun Shen and Yang Zhang},
title = {{Prompt Backdoors in Visual Prompt Learning}},
journal = {{CoRR abs/2310.07632}},
year = {2023}
}

@book{D02,
author = {Stephan Dempe},
title = {{Foundations of bilevel programming}},
publisher = {Springer Science \& Business Media},
year = {2002}
}

@misc{CIFAR,
title = {{CIFAR}},
howpublished = {\url{https://www.cs.toronto.edu/~kriz/cifar.html}},
}

@misc{SVHN,
howpublished = {\url{http://ufldl.stanford.edu/housenumbers/}},
}

@misc{Landing_AI,
howpublished = {\url{https://landing.ai/}},
}

@misc{frog,
howpublished = {\url{https://knowyourmeme.com/memes/pepe-the-frog}},
}

@misc{KCSS23,
title = {{Foundational vision models and visual prompt engineering for autonomous driving applications}},
author = {Gopi Krishnamurthy and Francisco Calderon Rodriguez and Shreyas Subramanian and Sujitha Martin},
year = {2023},
howpublished = {\url{https://aws.amazon.com/blogs/machine-learning/foundational-vision-models-and-visual-prompt-engineering-for-autonomous-driving-applications/}},
}
\end{small}

\end{document}